\documentclass[11pt,a4paper,pdftex]{article}

\usepackage[utf8]{inputenc}
\usepackage[T1]{fontenc}
\usepackage{amsmath}
\usepackage{amsthm}
\usepackage{amssymb}
\usepackage{xcolor}
\usepackage{enumerate}
\usepackage{booktabs}
\usepackage{fancyvrb}
\usepackage[maxnames=10, backend=biber]{biblatex}
\usepackage{hyperref}
\usepackage{orcidlink}

\newtheorem{thm}{Theorem}
\newtheorem{prop}[thm]{Proposition}
\newtheorem{lem}[thm]{Lemma}
\newtheorem{cor}[thm]{Corollary}

\theoremstyle{definition}
\newtheorem{defi}[thm]{Definition}

\newtheorem{rmk}[thm]{Remark}
\newtheorem{rmks}[thm]{Remarks}
\newtheorem{ex}[thm]{Example}
\newtheorem{algo}[thm]{Algorithm}

\DeclareMathAlphabet{\mymathbb}{U}{BOONDOX-ds}{m}{n}

\newcommand{\C}{ \mathbb C } % complexes
\newcommand{\R}{ \mathbb R } % réels
\newcommand{\Q}{ \mathbb Q } % rationnels
\newcommand{\Z}{ \mathbb Z } % entiers relatifs
\newcommand{\A}{ \mathbb A } % Adèles
\newcommand{\U}{ \mathbb U } % unités
\newcommand{\cL}{ \mathcal L } % fonctions intégrables
\newcommand{\ga}{ \mathfrak a } % idéal
\newcommand{\gm}{ \mathfrak m } % idéal
\newcommand{\gn}{ \mathfrak n } % idéal
\newcommand{\gf}{ \mathfrak f } % conducteur
\newcommand{\gp}{ \mathfrak p } % idéal premier
\newcommand{\gq}{ \mathfrak q } % idéal premier
\newcommand{\gr}{ \mathfrak r } % idéal premier
\let\ot=\leftarrow
\newcommand\AF{\A_F}

\newcommand\AFs{\AF^\times} % idèles
\newcommand{\prodvmreal}{\prod_{\sigma\in\gm_\infty}}
\newcommand{\prodvnotmreal}{\prod_{\substack{\sigma\notin\gm_\infty\\\sigma\text{ real}}}}
\newcommand{\prodvcomplex}{\prod_{\sigma\text{ complex}}}
\newcommand{\prodvreal}{\prod_{\sigma\text{ real}}}

\newcommand{\rr}[1]{\textcolor{red}{#1}}
\newcommand{\gD}{\mathfrak{D}}
\newcommand{\Calg}{\bigl(\widehat{C}_F\bigr)^{\mathrm{a.a.}}}
\newcommand{\Cmalg}{\bigl(\widehat{C}_\gm\bigr)^{\mathrm{a.a.}}}
\DeclareMathOperator{\Nm}{N}
\DeclareMathOperator{\Tr}{Tr}
\newcommand{\cc}{\circ}
\newcommand{\solen}{\mathbb{V}}
\newcommand{\rootsofone}[1]{w_{\scriptscriptstyle{#1}}}
\newcommand{\eps}{\varepsilon}

\newcommand{\set}[1]{ \left\{ #1\right\} } % ensemble dans des accolades
\newcommand{\isomto}{ {\mathop{\longrightarrow}^\sim} }
\newcommand{\card}[1]{ \#{ #1} } % cardinal
\newcommand{\abs}[1]{ \left\lvert #1\right\rvert } % module

\DeclareMathOperator{\sgn}{sgn} % signe
\renewcommand{\Re}{\operatorname{Re}} % partie réelle
\DeclareMathOperator{\im}{Im} % image
\DeclareMathOperator{\GL}{GL} % groupe linéaire
\DeclareMathOperator{\Id}{Id}

\DeclareMathOperator{\rank}{rk}

\DeclareMathOperator{\Aut}{Aut} % groupe des automorphismes
\DeclareMathOperator{\End}{End} % Endomorphismes
\DeclareMathOperator{\Gal}{Gal} % Groupe de Galois
\DeclareMathOperator{\Hom}{Hom} % Homomorphismes
\DeclareMathOperator{\Cl}{Cl} % Groupe des classes
\DeclareMathOperator{\decomp}{\mathcal{D}}

\newcommand{\hi}[1]{%
   \ooalign{\hss\makebox[0pt]{\colorbox{red!20}{$#1$}}\hss\cr\phantom{$#1$}}%
}

\newcommand\kbd[1]{{\ttfamily #1}}
\DefineVerbatimEnvironment{code}{Verbatim}%
{%gobble=2,
 frame=lines,
 framerule=1.5pt}
\newcommand\lmfdbref[2]{\href{https://www.lmfdb.org/#2}{\texttt{#1}}}

\title{Computing groups of Hecke characters}
\author{Pascal Molin\footnote{Université Paris Cité and Sorbonne Université,
CNRS, INRIA, IMJ-PRG, F75013 Paris, France, pascal.molin@imj-prg.fr \orcidlink{0000-0003-2462-8751}}
~and Aurel Page\footnote{INRIA, Univ. Bordeaux, CNRS, Bordeaux INP, IMB,
UMR5251, F-33400 Talence, France, aurel.page@inria.fr \orcidlink{0000-0002-1247-4970}}}
\newif\ifarxiv\arxivtrue

\begin{document}

\maketitle

\begin{abstract}
  We describe algorithms to represent and compute groups of Hecke characters.
  We make use of an idèlic point of view and obtain the whole
  family of such characters, including transcendental ones.
  We also show how to isolate the algebraic characters, which are of particular
  interest in number theory.
  This work has been implemented in Pari/GP, and we illustrate our work with a
  variety of explicit examples using our implementation.
\end{abstract}

%\tableofcontents

\section{Introduction}

Hecke characters are, from the modern point of view, continuous characters of
id\`ele class groups, in other words automorphic forms for~$\GL_1$.
They were introduced by Hecke~\cite{HeckeWork} who proved the functional
equation of their $L$-function, and are the starting point of many developments
that blossom in modern number theory: automorphic $L$-functions via Tate's
thesis~\cite{TateThesis},
$\ell$-adic Galois representations via Weil's notion of algebraic
characters~\cite{Weil1956},
Shimura varieties via CM theory~\cite{TaniyamaCM}, and the
Langlands programme via class field theory and the global Weil
group~\cite{WeilGroup}.
Despite their fundamental role, Hecke characters have not received a full
algorithmic treatment, perhaps due to the fact that they are considered
well-understood compared to automorphic forms on higher rank groups.
The existing literature only describes how to compute with finite order
characters, since they are characters of ray class groups~\cite{CohenAdvanced},
and algebraic Hecke characters~\cite{WatkinsComputingHecke}.
As part of a collective effort to enumerate and compute $L$-functions,
automorphic representations and Galois representations, we believe that
the~$\GL_1$ case also deserves close scrutiny, and this is the goal of the
present paper.

We describe algorithms to compute, given a number field~$F$ and a modulus~$\gm$
over~$F$, a basis of the group of Hecke quasi-characters of modulus~$\gm$
(Algorithm~\ref{algo:group}) and its subgroup of
algebraic characters (Algorithm~\ref{algo:algchar}), in a form suitable for evaluation at arbitrary ideals and
decomposition into local characters (Algorithm~\ref{algo:log}).
In particular, we describe a polynomial time algorithm to compute the maximal CM
subfield of~$F$ (Algorithm~\ref{algo:cmsubfield}).
It is sometimes believed that the ad\`elic
point of view is not suitable for computational purposes; we claim the contrary,
and adopt an ad\`elic setting %point of view
throughout the paper.
Our implementation \cite{parigp:gchar}
in Pari/GP~\cite{PariGP} is available from version~2.15 of the software.
We provide examples that illustrate the use of our algorithms and showcase some
interesting features of Hecke characters: a presentation of the software
interface, small degree examples, illustrations of automorphic induction from
quadratic fields, examples of CM abelian varieties with emphasis on the rigorous
identification of the corresponding Hecke character, illustration of the density
of the gamma shifts of Hecke $L$-functions in the conjectured space of possible
ones (Proposition~\ref{prop:density}), examples of provably partially algebraic
Hecke characters (Proposition~\ref{prop:partcm}) and of twists of
$L$-functions by Hecke characters.

The only previous work
on computation of infinite order Hecke characters is that of
Watkins~\cite{WatkinsComputingHecke}, so we give a short comparison: in
Watkins's paper, only
algebraic characters were considered, and only over a CM field, whereas we treat
arbitrary Hecke characters over arbitrary number fields; the values of
characters were represented exactly by algebraic numbers, whereas we represent
values by approximations since this is forced in the transcendental case; the
emphasis was on individual Hecke characters, which the user had to construct by
hand, whereas our emphasis is on groups of Hecke characters, which we construct
for the user, simply from the modulus.

Our implementation makes it possible to tabulate Hecke characters and their
$L$-functions systematically by increasing analytic conductor; we think that this
is a valuable project but we leave it for future work.

The paper is organized as follows. In Section~\ref{sec:reminder} we recall the
definitions and basic properties of Hecke characters and their $L$-functions.
In Section~\ref{sec:computing} we describe our algorithms to compute groups of
Hecke characters and evaluate them. In Section~\ref{sec:algebraic} we
present our algorithms to compute the maximal CM subfield and groups of
algebraic Hecke characters. Finally, Section~\ref{sec:examples} contains a
variety of examples.

\paragraph{Acknowledgements}
We thank the anonymous reviewers for their careful reading of our manuscript and their
many comments and suggestions. We also thank Karim Belabas and Bill Allombert for their
help in integrating our code to Pari/GP. The first author acknowledges support
of ANR FLAIR ANR-17-CE40-0012. The second author was supported by the grants ANR
CIAO ANR-19-CE48-0008 and ANR CHARM ANR-21-CE94-0003.

\section{Hecke characters}\label{sec:reminder}

We recall the definition of Hecke characters in the adèlic setting.
This material is standard and can be found in
\cite[chap. XIV]{LangANT} or \cite{Schappacher1988}.

Let $F$ be a number field of degree $[F:\Q]=n$ and discriminant~$\Delta_F$.
When~$K/F$ is a finite extension, we denote by~$\Nm_{K/F}$ the norm from~$K$ to~$F$;
we also denote~$\Nm = \Nm_{F/\Q}$ when~$F$ is clear from the context.
%Let~$\zeta$ be a generator of the group of roots of unity in~$F$.
For every prime ideal~$\gp$ of~$F$, we consider the completion~$F_\gp$
and its ring of integers $\Z_\gp$.
We choose a uniformizer~$\pi_\gp\in\Z_\gp$ and
denote by~$v_\gp \colon F_\gp^\times \twoheadrightarrow \Z$ the~$\gp$-adic valuation.
We will always use~$\sigma$ to denote an archimedean place of~$F$ and the
corresponding real or complex embedding.
For every place~$v$, let~$n_v = [F_v:\Q_v]$, and let~$|\cdot|_v$ be the normalized
absolute value, i.e.~$n_\sigma=1$ and~$ |\cdot |_\sigma = |\cdot |$ for a real
embedding~$\sigma$, $n_\sigma=2$ and~$|\cdot|_\sigma = |\cdot|^2$ for a complex
embedding~$\sigma$, and $|\pi_\gp|_\gp = \Nm(\gp)^{-1}$ for a prime ideal~$\gp$.
%Let~$m$ be the order of~$\mu_\infty(F)$.
We denote by $\AFs = \prod'F_v^\times$
the group of idèles of $F$. We write~$F_\R = F\otimes_\Q\R \cong
\prod_{\sigma}F_\sigma \cong \R^{r_1}\times\C^{r_2}$, where~$r_1$ (resp.~$r_2$)
is the number of real embeddings (resp. pairs of non-real complex embeddings)
of~$F$.

Let~$\U$ denote the group of complex numbers of absolute value~$1$.
For~$G$ a topological group, $G^\cc$ will denote the connected component of~$1$
in~$G$.

\subsection{Pontryagin duality}\label{sec:pontryagin}

We recall some definitions and properties of locally compact abelian groups
that will be used later.
See~\cite{MorrisDualityLayman, MorrisDuality} for general reference.

Let~$G$ be a locally compact abelian group.
A \emph{quasi-character} of~$G$ is a continuous morphism
\[
  \chi\colon G \to \C^\times.
\]

A \emph{character} of~$G$ is a continuous morphism
\[
  \chi\colon G \to \U.
\]

The group of characters of~$G$, which we denote by~$\widehat{G}$, is the Pontryagin
dual~$\Hom_{\rm cont}(G,\U)$ of~$G$, and is a locally compact abelian group.
The canonical map
\[
  G \to \widehat{\widehat{G\, }} %https://tex.stackexchange.com/questions/54712/how-does-one-make-a-double-widehat
\]
given by~$g \mapsto (\chi \mapsto \chi(g))$ is an isomorphism.
Let~$H\subset G$ be a subgroup. Let
\[
  H^\perp = \{\chi\in \widehat{G} \mid \chi(h) = 1 \text{ for all }h \in H\}
\]
be the Pontryagin orthogonal of $H$ in $\widehat G$.
Then~$H^\perp$ is a closed subgroup of~$\widehat{G}$, and $(H^\perp)^\perp$ is
the closure of~$H$, where the second orthogonal is taken in~$G$. If~$H$ is a
closed subgroup of~$G$, then we have canonical isomorphisms
\[
  \widehat{G/H} \cong H^\perp \text{ and } \widehat{G}/(H^\perp) \cong
  \widehat{H}.
\]

The group~$G$ is compact if and only if~$\widehat{G}$ is discrete.
%G profinite iff Ghat discrete torsion
%G compact connected iff Ghat discrete torsion-free

Pontryagin duality is an exact contravariant functor on the category of locally
compact abelian groups.

Let~$(x,y) \mapsto x\cdot y$ denote a nondegenerate $\R$-bilinear form on a
finite dimensional $\R$-vector space~$V$.
The pairing~$V\times V \to \U$ defined by~$(x,y) \mapsto \exp(2i\pi x\cdot y )$
induces an isomorphism~$V \cong \widehat{V}$.
We will use this isomorphism to identify characters on~$V$ with elements of $V$.

Let~$\Lambda$ be a full rank lattice in $V$.
The pairing above identifies the dual lattice~$\Lambda^\vee = \Hom(\Lambda,\Z)$ with the subgroup
\[
  \Lambda^\perp = \{x\in V \mid x\cdot y \in \Z \text{ for all }y \in \Lambda\},
\]
which is canonically isomorphic to~$\widehat{V/\Lambda}$ by the above,
and we have $\widehat{\Lambda} \cong V/\Lambda^\perp$.
In particular for $V = \R$ and $\Lambda = \Z$ we consider the standard
bilinear form and we have
$\widehat{\R/\Z}=\Z^\perp=\Z$ and $\widehat{\Z}=\R/\Z$.

The dual~$\solen = \widehat{\Q}$ of the group of rationals equipped with the
discrete topology, is the compact topological group~$\displaystyle \lim_{\ot n}\R/n\Z$, called
the \emph{solenoid}.

\subsection{General Hecke characters}

A Hecke quasi-character is a quasi-character of~$C_F = \AFs/F^\times$,
and a Hecke character is a character of~$C_F$.

The \emph{norm} is the Hecke quasi-character
\[
  \|\cdot\| \colon C_F \to \C^\times
\]
defined by
\[
  x = (x_v)_v \mapsto \|x\| = \prod_v |x_v|_v.
\]
This is a well-defined Hecke quasi-character by the product formula.

Every Hecke quasi-character~$\chi$ is of the form~$\chi = \chi_0 \|\cdot\|^s$
for a unique Hecke character~$\chi_0$ and a unique~$s\in\R$.
We refer to~$\chi_0$ as the \emph{unitary component} of $\chi$.
In the algebraic setting, the value $w=-2s$ is the \emph{weight} of $\chi$.

We also define~$C_F^1 = \ker (\|\cdot\|\colon C_F \to \R_{>0})$ to be the kernel
of the norm, which is a compact group.
We have a canonical embedding
\[
  \R_{>0} \to C_F,
\]
by sending~$t\in\R_{>0} \mapsto ((t^{1/n})_\sigma,1,\dots)\in\AFs$
where~$t\mapsto (t^{1/n})_\sigma$ denotes the diagonal embedding~$\R_{>0} \to
\prod_\sigma F_\sigma^\times$, and a
canonical decomposition
\[
  C_F \cong C_F^1 \times\R_{>0}.
\]

As a consequence, it suffices to compute the characters
of $C_F^1$ to deduce the full groups of Hecke characters and
Hecke quasi-characters
\begin{equation}
  \label{eq:normsplitting}
    \Hom_{\rm cont}(C_F,\C^\times)
    = \widehat{C}_F \|\cdot\|^\R
    = \widehat{C}^1_F\|\cdot\|^\C.
\end{equation}

Every quasi-character~$\chi$ of~$\AFs$ (and in particular every Hecke quasi-character)
admits a factorization~$\chi = \prod_v\chi_v$, where~$\chi_v$ is a
quasi-character of~$F_v^\times$. We therefore describe quasi-characters of local
fields.

\subsection{Local characters}

\begin{itemize}
    \item
Every quasi-character~$\chi$ of~$\C^\times$ is of the form
\[
  \chi(z) = \Bigl(\frac{z}{|z|}\Bigr)^k |z|_\C^s = \Bigl(\frac{z}{|z|}\Bigr)^k |z|^{2s}
\]
for a unique pair~$(k,s)\in\Z\times\C$.
The quasi-character~$\chi$ is a character if and only if~$\Re(s) = 0$, i.e.~$s = i\varphi$ for
some~$\varphi\in\R$.

\item
Every quasi-character~$\chi$ of~$\R^\times$ is of the form
\[
  \chi(x) = \sgn(x)^k |x|^s
\]
for a unique pair~$(k,s)\in\set{0,1}\times\C$.
%We will often choose~$k\in\{0,1\}$.
We say that~$\chi$ is \emph{unramified} if~$k=0$.
The quasi-character~$\chi$ is a character if and only if~$\Re(s) = 0$, i.e.~$s = i\varphi$ for
some~$\varphi\in\R$.

\item
%Let~$\pi_\gp$ be a uniformizer of~$F_\gp$.
Let~$\gp$ be a prime ideal of~$\Z_F$.
Every quasi-character~$\chi$ of~$F_\gp^\times$ is of the form
\[
  \chi(x) = \chi_0(x\pi_\gp^{-v_\gp(x)} \bmod \gp^m) \chi(\gp)^{v_\gp(x)}
\]
for a unique~$m\ge 0$ and a unique primitive character~$\chi_0$
of~$(\Z_\gp/\gp^m)^\times$, and where we
write~$\chi(\gp)=\chi(\pi_\gp)\in\C^\times$.
Note that in general~$\chi(\gp)$
depends on the choice of uniformizer~$\pi_\gp$, but~$\chi(\gp)$ is well defined up to the
roots of unity of the same order as~$\chi_0$.
We call~$\gp^m$ the conductor of~$\chi$ and~$m$ its conductor exponent.
If~$m=0$ we call~$\chi$ \emph{unramified}; in this case, $\chi(\gp)$ does not
depend on the choice of uniformizer, and the quasi-character~$\chi$ only depends
on~$\chi(\gp)$.
Regardless of~$m$, the quasi-character~$\chi$ is a character if and only
if~$\chi(\gp)\in\U$.
\end{itemize}

Whenever we write a global id\`ele character~$\chi$ as a product of local
characters~$\chi_v$, we write its local parameters~$k_\sigma,\varphi_\sigma$,
and~$m_\gp$, and we let~$\gf_\chi = \prod_\gp \gp^{m_\gp}$ be the conductor of~$\chi$.
Note that for a complex place, the pair~$(k_\sigma,\varphi_\sigma)$ depends on the choice
of a complex embedding among the two conjugate ones, or equivalently on the
choice of an isomorphism between the completion of~$F$ and~$\C$: we
have~$\varphi_{\bar{\sigma}} = \varphi_{\sigma}$ and~$k_{\bar{\sigma}} =
-k_{\sigma}$.

\subsection{$L$-function}
Let~$\chi$ be a Hecke character such that~$\sum_\sigma n_\sigma\varphi_\sigma =
0$, i.e. that is trivial on the embedded~$\R_{>0}$ in~\eqref{eq:normsplitting}.
Let~$N_\chi = |\Delta_F|\cdot \Nm(\gf_\chi)$.
Let
\[
  L(\chi,s) = \prod_{\gp\nmid \gf_\chi}(1-\chi(\gp)\Nm(\gp)^{-s})^{-1}
\]
and
\[
  \gamma(\chi,s) =
  \prodvreal\Gamma_{\R}(s+i\varphi_\sigma+k_\sigma)
  \cdot
  \prodvcomplex\Gamma_{\C}(s+i\varphi_\sigma+|k_\sigma|/2).
\]
where $\Gamma_\R(s)=\pi^{-\frac s2}\Gamma(\frac s2)$ and
$\Gamma_\C(s)=2(2\pi)^{-s}\Gamma(s)$. Then
\[
  \Lambda(\chi,s) = N_\chi^{s/2}\gamma(\chi,s)L(\chi,s)
\]
satisfies the functional equation
\[
  %\Lambda(\bar{\chi},1-s) = W(\chi) \Lambda(\chi,s)
  \Lambda(\chi,1-s) = W(\chi) \Lambda(\bar{\chi},s)
\]
for some complex number~$W(\chi)$ of absolute value~$1$.

\ifarxiv
We have the formula
%\[
%  W(\chi) = \prodvcomplex (4^{i\varphi_\sigma}i^{|k_\sigma|})
%  \cdot \prodvreal i^{k_\sigma}
%  \cdot (\Nm \gf_\chi)^{-1/2}
%  \cdot \prod_{\gp\mid\gf_\chi} \bigl(\chi(\gp)^{d_\gp}
%  \overline{\tau(\chi_\gp)}\bigr)
%  \cdot \prod_{\gp\nmid\gf_\chi}\chi(\gp)^{d_\gp},
%\]

\[
  W(\chi) = \prod_v W(\chi_v),
\]
where
\[
  W(\chi_v) =
  \begin{cases}
    4^{i\varphi_\sigma}i^{|k_\sigma|}  \text{ if }v=\sigma\text{ is complex,}\\
    i^{k_\sigma}  \text{ if }v=\sigma\text{ is real,}\\
    \chi(\gp)^{d_\gp} \overline{\tau(\chi_\gp)} \Nm(\gp)^{-m_\gp/2} \text{ if
    }v=\gp\mid \gf_\chi\text{, and} \\
    \chi(\gp)^{d_\gp}\text{ if }v=\gp\nmid\gf_\chi.
  \end{cases}
\]
where~$d_\gp = v_\gp(\gD \gf_\chi)$ and~$\gD$ is the different of~$F$ (so that the product is finite),
and
\[
  \tau(\chi_\gp) =
  \sum_{\epsilon\in(\Z_\gp/\gf_{\chi,\gp})^\times}\chi(\epsilon)\exp(2i\pi
  \lambda\circ \Tr_{F_\gp/\Q}(\epsilon/\pi_\gp^{d_\gp}))
\]
where~$\lambda\colon \Q_p \to \Q_p/\Z_p \to \Q/\Z$.
\fi

\subsection{Algebraic Hecke characters}

\textbf{Warning}: an algebraic Hecke character is usually not a Hecke character,
it is only a quasi-character.

Let~$\chi$ be a Hecke quasi-character. It is called~\emph{algebraic} if for
every archimedean place~$\sigma$ of~$F$, there exists integers~$p_\sigma,q_\sigma\in
\Z$ such that for all~$z\in (F_\sigma^\times)^\cc$ we have \footnote{The choice
of sign in the exponents is such that the values of~$\chi$ at integral ideals
are algebraic integers if and only if all~$p_\sigma$ and~$q_\sigma$ are
nonnegative.}
\[
  \chi_\sigma(z) = z^{-p_\sigma}(\bar{z})^{-q_\sigma}.
\]

Note: if~$\sigma$ is complex, then~$p_\sigma$ and~$q_\sigma$ are uniquely
determined; if~$\sigma$ is real then only their sum is well-defined.
We say that~$\chi$ is \emph{of type}~$(p_\sigma,q_\sigma)_\sigma$.

\begin{ex}
  The norm~$\|\cdot\|$ is an algebraic character,
  of type~$(p_\sigma,q_\sigma)=(-1,-1)$ if~$\sigma$ is complex.
  We have~$\|\gp\| = \Nm(\gp)^{-1}$ for every prime ideal~$\gp$.
\end{ex}

\begin{defi}\label{defi:algebraic}
We call a Hecke character \emph{almost-algebraic} if~$\varphi_\sigma=0$ for all~$\sigma$.
We denote by $\Calg$ the subgroup of almost-algebraic characters.
\end{defi}

\begin{rmk}
Algebraic characters correspond to~\emph{type~$A_0$} and almost-algebraic
to~\emph{type~$A$} with trivial norm component in Weil's terminology \cite{Weil1956}.
By a theorem of Waldschmidt \cite{Waldschmidt1981}, these definitions coincide with the fact
that a quasi-character has~type~$A$ if and only if its values are algebraic, and~type~$A_0$
if and only if there exists a finite extension of~$\Q$
containing all of its values.
\end{rmk}

\subsubsection{Parameters at infinity of algebraic Hecke characters}
\label{sec:almostalg}
It is known that if $F$ has a real embedding, then every algebraic Hecke character is an
integral power of the norm times a Hecke character of finite order
(see~\cite{Weil1956}). So from now on we
assume that~$F$ is totally complex.
We recall the following well-known lemma.

\begin{lem}
  \label{lem:infinitytype}
  Let~$\chi_0$ be a Hecke character and let~$(k_\sigma,\varphi_\sigma)$ denote its
  local parameters at infinite places.
  The character~$\chi_0$ is the unitary component of an algebraic Hecke
  character if and only if~$\chi_0$ is almost algebraic and all~$k_\sigma$ have
  the same parity.

  More precisely, let~$\chi = \chi_0\|\cdot\|^{-w/2}$ be a Hecke quasi-character
  with~$w\in\R$. If~$\chi$ is algebraic of type~$(p_\sigma,q_\sigma)$, then
  \begin{itemize}
    \item $w\in\Z$;
    \item $p_\sigma+q_\sigma=w$ for all~$\sigma$;
    \item $k_\sigma = q_\sigma-p_\sigma$ for all~$\sigma$;
    \item $\varphi_\sigma = 0$ for all~$\sigma$.
  \end{itemize}
  Conversely, if~$\chi_0$ is almost-algebraic and all~$k_\sigma$ have the same
  parity, let~$w\in\Z$ have the same parity as the~$k_\sigma$; then~$\chi =
  \chi_0\|\cdot\|^{-w/2}$ is algebraic.
\end{lem}
\begin{proof}
  Let~$\chi = \chi_0\|\cdot\|^{-w/2}$ be a Hecke quasi-character
  with~$w\in\R$, so that for all~$z\in\C^\times$ we have
  \[
    \chi_\sigma(z) =
  \Bigl(\frac{z}{|z|}\Bigr)^{k_\sigma}|z|^{2i\varphi_\sigma-w}.
  \]
  Let~$p,q\in\Z$. For all~$z\in\C^\times$ we have
  \[
    z^{-p}(\bar{z})^{-q} = 
  \Bigl(\frac{z}{|z|}\Bigr)^{q-p}|z|^{-p-q}.
  \]
  By uniqueness of parameters of quasi-characters of~$\C^\times$, the
  quasi-character~$\chi$ is algebraic of type~$(p_\sigma,q_\sigma)$ if and only
  if for all~$\sigma$ we have $k_\sigma = q_\sigma-p_\sigma$, $\varphi_\sigma=0$
  and $w = p_\sigma+q_\sigma$.
  In this case, $\chi_0$ is almost-algebraic and for all~$\sigma$ we
  have~$k_\sigma \equiv q_\sigma-p_\sigma \equiv p_\sigma+q_\sigma \equiv w \bmod{2}$, so
  that all~$k_\sigma$ have the same parity. This also validates the construction of
  an algebraic~$\chi$ from an almost-algebraic~$\chi_0$ satisfying the parity
  condition.
\end{proof}

%\begin{lem}
%    %\label{lem:infinitytype}
%Let~$\chi_0$ be a Hecke character such that all~$\varphi_\sigma$ are $0$
%and all the~$k_\sigma$ have the same parity.
%
%Let~$w\in\Z$ be such that~$k_\sigma+w$ is even for all~$\sigma$.
%Then~$\chi = \chi_0\|\cdot\|^{-\frac w2}$ is an algebraic Hecke character of
%infinity-type~$((p_\sigma,q_\sigma))_\sigma$,
%    where~$p_\sigma = \frac{w-k_\sigma}{2}\in\Z$
%    and~$q_\sigma = \frac{w+k_\sigma}{2}\in\Z$.
%
%    Conversely, an algebraic character of infinity-type $((p_\sigma,q_\sigma))_\sigma$
%    must have constant value $p_\sigma+q_\sigma=w$, in which case its parameters
%    are $k_\sigma=q_\sigma-p_\sigma$.
%\end{lem}
%\begin{proof}
%Assume~$\chi = \chi_0\|\cdot\|^s$ is an algebraic character, with~$\chi_0$ a
%Hecke character and~$s\in\R$, and let~$\chi_\sigma$ denote the local component
%of~$\chi_0$ at a complex place~$\sigma$. We solve
%\[
%    \chi_\sigma(z)|z|_\sigma^{s} =
%  \Bigl(\frac{z}{|z|}\Bigr)^{k_\sigma}|z|^{2(i\varphi_\sigma+s)}
%  = z^{k_\sigma/2+i\varphi_\sigma+s}(\bar{z})^{-k_\sigma/2+i\varphi_\sigma+s}
%  = z^{-p_\sigma}(\bar{z})^{-q_\sigma}.
%\]
%as
%\[
%  p_\sigma = -k_\sigma/2-i\varphi_\sigma-s \text{ and }q_\sigma =
%  k_\sigma/2-i\varphi_\sigma-s.
%\]
%This implies that~$\varphi_\sigma = 0$ and that~$2s\in\Z$, so we write~$w = -2s$.
%We obtain
%\[
%  2p_\sigma = -k_\sigma + w,\
%  2q_\sigma = k_\sigma + w,\
%  w = p_\sigma + q_\sigma\text{, and }
%  k_\sigma = q_\sigma - p_\sigma,
%\]
%so all the~$k_\sigma$ must have the same parity.
%\end{proof}

Thus the group of unitary components~$\chi_0$ of algebraic Hecke
characters~$\chi=\chi_0\|\cdot\|^{-w/2}$ is a
finite index subgroup of the group of almost-algebraic Hecke characters.

\subsubsection{$L$-function of an algebraic Hecke character}
Let~$\chi = \chi_0\|\cdot\|^{-w/2}$ be an algebraic Hecke character as above.
Let~$\gf_\chi = \gf_{\chi_0}$ be its conductor and~$N_\chi = N_{\chi_0}$.
Let
\[
  L(\chi,s) = \prod_{\gp\nmid \gf_\chi}(1-\chi(\gp)\Nm(\gp)^{-s})^{-1}
  = L(\chi_0,s-w/2),
\]
and
\[
  \gamma(\chi,s) = \prod_\sigma \Gamma_{\C}(s-\min(p_\sigma,q_\sigma))
  = \gamma(\chi_0,s-w/2).
\]
Then
\[
  \Lambda(\chi,s) = N_\chi^{s/2}\gamma(\chi,s)L(\chi,s)
\]
satisfies the functional equation
\[
  \Lambda(\chi, w+1-s) = W(\chi) \Lambda(\bar{\chi},s)
\]
for some complex number~$W(\chi) = W(\chi_0)$ of absolute value~$1$.

\section{Computing the group of Hecke characters}\label{sec:computing}

\subsection{Filtration by modulus}
\label{sec:filtration}
We have a non-canonical isomorphism
\[
    \widehat{C}_F \cong T \times \Q^{r_1+r_2-1} \times \Z^{r_2} \times \R,
\]
where~$T$ is an infinite torsion abelian group. Indeed, we have the classical
decomposition~\cite{WeilGroup}
\[
  1 \to C_F^\cc \to C_F \to \pi_0(C_F) \to 1 \text{, where }C_F^\cc\cong
  \solen^{r_1+r_2-1} \times (\R/\Z)^{r_2}\times \R ,
\]
where~$\solen = \widehat{\Q}$ is the solenoid, and~$\pi_0(C_F)$ is
profinite; by Pontryagin duality, we get
\[
    0 \to T \to \widehat{C}_F \to \Q^{r_1+r_2-1}\times \Z^{r_2} \times \R\to 0,
\]
and this exact sequence splits. % since~$\R\times\Q^{r_1+r_2-1}\times \Z^{r_2}$ is
%injective as a~$\Z$-module. this does not prove that the sequence splits topologically!
Since we cannot give a finite description of the whole
group~$T$, we will filter~$\widehat{C}_F$ according to moduli.

Let $\gm = \gm_f\gm_\infty$ be a modulus, meaning that~$\gm_f$ is an integral ideal
and~$\gm_\infty$ is a set of real embeddings of~$F$.
We write
\[
  (\Z_F/\gm)^\times = (\Z_F/\gm_f)^\times \times \prodvmreal\set{\pm 1}.
\]

%A Hecke character is a continuous morphism
%\[ \chi : \AFs/F^\times \to \U.\]
A Hecke character~$\chi$ is said to have modulus $\gm$ if
$\chi$ is trivial on the group~$U(\gm)$ of idèles congruent to 1 mod $\gm$:
\[
    U(\gm) =
    \prod_{\gp\mid \gm_f}(1+\gp^{v_\gp(\gm)}\Z_\gp)
    \times\prod_{\gp\nmid\gm_f}\Z_\gp^\times
    \times\prodvmreal\set{1}
    \times\prodvnotmreal\set{\pm 1}
    \times\prodvcomplex\set{1}.
\]
Equivalently, the conductor of~$\chi$ divides~$\gm_f$ and~$\chi$ is unramified
at all the real places not dividing~$\gm_\infty$.

The character group of modulus $\gm$ is the dual of
\[ C_\gm = \AFs/(F^\times \cdot U(\gm)), \]
and we have
\[
  \widehat{C}_F = \bigcup_{\gm}\widehat{C}_\gm.
\]

In the remainder of this section, we fix a modulus~$\gm$.

\subsection{Explicit description}

The character group~$\widehat{C}_\gm$ is isomorphic
to~$T_\gm \times \Z^{n-1} \times \R$ where~$T_\gm$ is finite.
Our goal in the next paragraphs is to prove the following
\begin{prop}
    \label{prop:explicit}
There exist an integer $\ell\geq 0$, a lattice~$\Lambda$ %\in\Z^k\times\R^n$
    of rank~$\ell+n-1$,
and two isomorphisms
\[
    \begin{aligned}
        \cL\colon &C_\gm \isomto (\Z^\ell\times\R^n)/\Lambda \\
        \cL^*\colon &\widehat{C}_\gm \isomto \Lambda^\bot/\Z^\ell % \subset (\R/\Z)^k\times\R^n
    \end{aligned}
\]
where $\Lambda^\bot$ is the Pontryagin orthogonal of $\Lambda$ in $\R^{\ell+n}$,
and
such that for all $\chi\in \widehat{C}_\gm$ and $x\in\AFs$
we have
\begin{equation}
  \label{eq:chareval}
  \chi(x) = \exp(2i\pi \cL^*(\chi)\cdot \cL(x)).
\end{equation}
\end{prop}
The lattice~$\Lambda$ and the isomorphisms~$\cL$ and~$\cL^*$ will be made
explicit in the next subsections.

\subsection{Idèle class groups}

\begin{defi}
  Let~$x\in\A_F^\times$. We define the \emph{ideal attached to~$x$} to be
  \[
    \prod_{\gp}\gp^{v_\gp(x_\gp)}.
  \]
  Let~$S$ be a finite set of primes of~$F$.
  Define the group of \emph{$S$-idèles} to be
  \[
    U_S =
          \prod_{\gp\in S}F_\gp^\times
          \times \prod_{\gp\notin S}\Z_\gp^\times
          \times F_\R^\times,
  \]
  and the group of~\emph{$S$-units} $\Z_{F,S}^\times = F^\times \cap U_S$.
\end{defi}

\begin{lem}
    \label{lem:surj2}
  Let~$x\in\A_F^\times$.
  Then~$x\in U_S$ if and only if the ideal attached to~$x$ belongs to the
  group~$\langle S\rangle$ generated by~$S$.
  If~$S$ generates the class group of~$F$, then~$\A_F^\times = U_S \cdot F^\times$.
\end{lem}
\begin{proof}
  The first property follows from rewriting the definition of~$U_S$ as~$U_S =
  \{x\in \A_F^\times \mid v_\gp(x) = 0 \text{ for all }\gp\notin S\}$.
  Let~$x\in \A_F^\times$ and~$\ga$ the ideal attached to~$x$.
  Assuming~$S$ generates the class group, let~$\alpha$ be such
    that~$\ga(\alpha^{-1})\in \langle S\rangle$. Then~$x\alpha^{-1}\in U_S$.
\end{proof}

\newcommand{\GSm}{\Z^S \times (\Z_F/\gm)^\times \times (F_\R^\times)^\cc}
\begin{defi}\label{def:decomp}
  Let~$S$ be a set of primes generating the class group of~$F$.
  Let
  \[
    \decomp_S \colon U_S \to \GSm
  \]
  be defined by
  \[
    \decomp_S(x) = \bigl(
    v_\gp(x_\gp)_{\gp\in S},
    (u\bmod \gm_f),
    (\sgn(x_\sigma))_{\sigma\in\gm_\infty},
    (|x_\sigma|)_{\sigma\text{ real}},
    (x_\sigma)_{\sigma\text{ complex}}
    \bigr),
  \]
  where~$u\in\prod_\gp\Z_\gp^\times$ is defined by~$u_\gp = x_\gp
  \pi_\gp^{-v_\gp(x_\gp)}$,
  where we recall that~$\pi_\gp\in\Z_\gp$ is a chosen uniformiser.

  Let
  \[
    \decomp \colon \A_F^\times/F^\times \to \Bigl[\GSm\Bigr] / \decomp_S(\Z_{F,S}^\times)
  \]
  be defined by~$\decomp(x\cdot F^\times) = \decomp_S(x\alpha^{-1})$ where~$\alpha\in F^\times$
  such that~$x\alpha^{-1}\in U_S$.
\end{defi}

\begin{lem}
\label{lem:cmisom2}
Let $S$ be a finite set of primes generating the class group.
Then~$\decomp$ is well-defined and induces an isomorphism
\begin{equation}
    \label{eq:finalquotient2}
  C_\gm \cong
  \Bigl[\GSm\Bigr]/\decomp_S(\Z_{F,S}^\times).
\end{equation}
\end{lem}
\begin{proof}
  The existence of the element~$\alpha$ from the definition of~$\decomp$ exists
  by Lemma~\ref{lem:surj2}.
  If~$x\alpha^{-1}$ and~$x\beta^{-1}$ belong to~$U_S$ with~$\alpha,\beta\in
  F^\times$, then~$\beta/\alpha\in F^\times\cap U_S = \Z_{F,S}^\times$,
  so~$\decomp$ is well-defined.
  By the decompositions~$F_\gp^\times \cong \pi_\gp^\Z \times \Z_\gp^\times$
  and~$F_\R^\times \cong \{\pm 1\}^{r_1}\times (F_\R^\times)^\circ$ and the
  Chinese remainder theorem, the map~$\decomp_S$ is onto, and~$\ker \decomp_S = U(\gm) \subset U_S$.
  Moreover by definition~$\decomp(F^\times) = 1$.
  This proves that~$\ker\decomp = F^\times\cdot U(\gm)$ and therefore~$\decomp$
  induces an isomorphism from~$C_\gm = \A_F^\times/(F^\times\cdot U(\gm))$ to
  its codomain.
\end{proof}

\subsection{Logarithm maps}

In this section we fix a finite set~$S$ of primes that generates the class group
of~$F$ and a modulus~$\gm$.

\begin{defi}
    \label{def:mapL}
Consider the usual archimedean
    logarithm $\log_\infty:(F_\R^\times)^\cc\to\R^{r_1+r_2}\times(\R/\Z)^{r_2} = \R^n/\Z^{r_2}$
\begin{equation}
    \label{eq:logoo}
    \log_\infty(z) =
    \left(\Bigl(\frac{n_\sigma}{2\pi}\log\abs{z_\sigma}\Bigr)_{\sigma},
    \Bigl(\frac{\arg(z_\sigma)}{2\pi}\Bigr)_{\sigma\text{ complex}}\right),
\end{equation}
  and choose an integer~$r(\gm)\ge 0$, a full sublattice~$\Lambda_{\gm} \subset
  \Z^{r(\gm)}$ and an isomorphism
\begin{equation}
    \label{eq:logm}
    \log_\gm : (\Z_F/\gm)^\times \isomto \Z^{r(\gm)}/\Lambda_{\gm}.
\end{equation}

  Let~$\ell = \card{S}+{r(\gm)}$, and let
  \[
    \cL_S \colon U_S \to \frac{\Z^\ell\times \R^n}{\Lambda_{\gm}+\Z^{r_2}}
  \]
  be the composition of~$\decomp_S$ with
  \[
    \Id_{\Z^S} \times \log_\gm \times \log_\infty.
  \]
We identify $\Lambda_{\gm}$ and $\Z^{r_2}$ with their embedding in $\Z^\ell\times\R^n$.

  Let
  \begin{equation}
    \label{deflambda}
    \Lambda =  \cL_S(\Z_{F,S}^\times) + \Lambda_{\gm} + \Z^{r_2},
  \end{equation}
  and let
  \[
    \cL \colon \A_F^\times/F^\times \to \frac{\Z^\ell\times \R^n}{\Lambda}
  \]
  be defined by~$\cL(x \cdot F^\times) = \cL_S(x\alpha^{-1})$ where~$\alpha\in F^\times$
  is such that~$x\alpha^{-1}\in U_S$.
\end{defi}

\begin{defi}
We define the dual logarithm~$\cL^* \colon \widehat{C}_\gm \to
(\R/\Z)^{\ell}\times \R^n$ by
\begin{equation}
  \label{duallogmap}
  \cL^*(\chi) = \!
  \left(
    \Bigl(\frac{\arg \chi(\gp)}{2\pi}\Bigr)_{\gp\in S},
    \Bigl(\frac{\arg \chi(\log_\gm^{-1}(g_i))}{2\pi}\Bigr)_{i=1}^{r(\gm)},
    (\varphi_\sigma)_\sigma,
    (k_\sigma)_{\sigma \text{ complex}}
    \right)
\end{equation}
where~$(g_i)_{i=1}^{r(\gm)}$ is the image in~$\Z^{r(\gm)}/\Lambda_{\gm}$ of the standard basis
of~$\Z^{r(\gm)}$ and~$\varphi_\sigma,k_\sigma$ are the parameters at infinity
of~$\chi$.
\end{defi}
Recall that we defined~$\chi(\gp) = \chi(\pi_\gp)$, so that~$\cL^*$ depends on
the choices of~$\pi_\gp$ for~$\gp\in S$.

We now prove Proposition \ref{prop:explicit} in the following precise form.
\begin{prop}
Let $\Lambda^\bot$ be the Pontryagin orthogonal of~$\Lambda$
in~$\R^{\ell+n}$.
The homomorphisms~$\cL$ and~$\cL^*$ induce isomorphisms
\[
    \cL\colon C_{\gm} \longrightarrow \frac{\Z^\ell\times\R^n}{\Lambda}
    \text{ and }
    \cL^*\colon \widehat{C}_\gm \longrightarrow \Lambda^\bot/\Z^\ell.
\]
Let $\chi\in \widehat{C}_\gm$ be a character of
modulus $\gm$ and let~$x\in\AFs$, then
\begin{equation}
  \label{eq:chareval-bis}
  \chi(x) = \exp(2i\pi \cL^*(\chi)\cdot \cL(x)),
\end{equation}
  where~$(w,v) \mapsto w\cdot v$ denotes the standard inner product
  on~$\R^{\ell+n}$.
\end{prop}
\begin{proof}
  The fact that~$\cL$ is well-defined and induces an isomorphism follows
  immediately from Lemma~\ref{lem:cmisom2}.
  Applying Pontryagin duality to the sequence
  \[
     0\to C_\gm \to \R^{\ell+n}/\Lambda \to (\R/\Z)^\ell\to 0
  \]
  gives~$\widehat{C}_\gm = \Lambda^\bot/\Z^\ell$.

  Let~$x\in\A_F^\times$ and write~$x = \alpha \cdot x\alpha^{-1}$ with~$\alpha\in F^\times$
  and~$x\alpha^{-1}\in U_S$ by Lemma~\ref{lem:surj2}, and let~$u$ be as in
  Definition~\ref{def:decomp}. We have
  \[
    x = \alpha \prod_{\gp\in S}\pi_\gp^{v_\gp(x_\gp\alpha^{-1})} \cdot u \cdot
    \prod_\sigma (x_\sigma\sigma(\alpha)^{-1}),
  \]
  and therefore
  \[
    \chi(x) = \chi(\alpha)
    \cdot \prod_{\gp\in S}\chi_\gp(\pi_\gp^{v_\gp(x_\gp\alpha^{-1})})
    \cdot \chi(u)
    \cdot \prod_\sigma \chi_\sigma(x_\sigma \sigma(\alpha)^{-1})
    ,
  \]
  where~$\chi(\alpha)=1$ and~$\chi(u) = \prod_{\gp\mid\gm_f}\chi_\gp(u_\gp \bmod
  \gp^{m_\gp})$.
  By definition the product of local character evaluations is $\exp(2i\pi \cL^*(\chi)\cdot\cL(x))$.
  This also proves that the image of~$\cL^*$ lies in~$\Lambda^\perp$ and
  that~$\cL^*$ induces an isomorphism as claimed.
\end{proof}

\begin{rmk}
  The lattice $\Lambda$ is not cocompact in $\R^{\ell+n}$, so that the
  Pontryagin orthogonal~$\Lambda^\perp$ is not discrete. In the next section we
  factor out the norm, so that the resulting lattice is cocompact and its
  Pontryagin orthogonal can be expressed as a dual lattice as in
  Section~\ref{sec:pontryagin}.
\end{rmk}

\subsection{Characters modulo the norm}

Let~$C_\gm^1 = C_F^1\cap C_\gm = \ker (C_\gm \to \R_{>0})$ be the kernel of
the norm, which is compact.
We have a canonical splitting inherited from
(\ref{eq:normsplitting})
\[
  C_\gm \cong C_\gm^1\times\R_{>0},
\]
and the corresponding decomposition
\[
  \widehat{C}_\gm \cong \widehat{C}^1_\gm \times \|\cdot\|^{i\R}
\]
where $\widehat{C}^1_\gm$ is a discrete finitely generated abelian group.

%Let~$E = \R^{n+k}$ equipped with its standard inner product,
%let~$v_0\in E$ be the
%vector having coordinate~$n_\sigma$ at the component corresponding to the
%infinite place~$\sigma$ and~$0$ elsewhere, let
%\[
%  H = \left\{v\in E\mid \sum_{\sigma} n_\sigma v_\sigma = 0\right\} = v_0^\perp.
%\]
%Let~$p_H^\perp$ be the orthogonal projection onto $H$. We define~$\Lambda_1 =
%p_H^\perp(\Lambda)$ and~$\cL_1 = p_H^\perp\circ \cL \colon C_\gm^1 \to H/\Lambda_1$.

\begin{prop}
    \label{prop:groupmodnorm}
    Let~$v_0\in\R^{\ell+n}$ be the vector having coordinate~$n_\sigma$
    at the components corresponding to~$\varphi_\sigma$ and~$0$ elsewhere, and
    $p_0:\R^{\ell+n}\to (\R v_0)^\perp$ the orthogonal projection.

    Then $p_0\circ \cL$ induces an isomorphism
    \[
        \widehat{C}^1_\gm \cong p_0(\Lambda)^\vee /\Z^\ell.
    \]
    % where
    %\[ \Lambda_1^\vee = \Hom(p(\Lambda),\Z) \cong \set{v\in H \mid v\cdot p(\Lambda) \subset \Z}
    %\supset \Z^k \times \set{0}^n,
    %\]
\end{prop}
\begin{proof}
    Let $H=(\R v_0)^\perp=\set{x \mid \sum n_\sigma x_\sigma=0}$, we have an exact sequence
    \[
        0\to C_\gm^1 \to H/p_0(\Lambda)\to(\R/\Z)^\ell\to 0,
    \]
    where $p_0(\Lambda)$ has full rank in $H$, so that we identify
    $p_0(\Lambda)^\perp = p_0(\Lambda)^\vee$ in the dual sequence.
%\[
%    0\ot \widehat{C}^1_\gm \ot \widehat{(H/\Lambda_1)}=\Lambda_1^\vee \ot \Z^k \ot 0
%\]
\end{proof}

\begin{rmk}
By an appropriate choice of basis of the lattice $\Lambda$,
we naturally obtain a structured basis of $C_\gm$
according to the filtration
\[
  \widehat{\Cl(\gm)} \subset \widehat{C}^1_\gm
  \subset \widehat{C}_\gm.
\]
It is even possible to obtain a basis exhibiting the filtration
\[
  \widehat{\Cl_F} \subset \widehat{\Cl(\gm)} \subset (\widehat{C}^1_\gm)_{k=0} \subset \widehat{C}^1_\gm
  \subset \widehat{C}_\gm,
\]
but our implementation makes a different choice of basis, using an SNF basis for
the torsion subgroup and exhibiting the subgroup of almost-algebraic characters,
  as explained in Section~\ref{sec:algebraic}.
\end{rmk}

\subsection{Algorithms}

Since a precise discussion of the complexity is not the main point of the paper,
we delegate the difficult operations to oracles.

\begin{defi}\label{def:bnfinit}
Let $F$ be a number field and $I_F$ the set of fractional ideals of $\Z_F$.
We say that $F$ is \emph{strongly computable} if it is equipped with
\begin{itemize}
    \item algorithms to compute field operations in $F$,
        factorizations into prime ideals and valuations in $I_F$;
    \item a finite set $S$ of prime ideals generating the class group;
    \item generators of the $S$-units $\Z_{F,S}^\times$;
    \item a principalization oracle $p_S\colon I_F\to F^\times\times\Z^S$ such
      that for every ideal~$\ga\in I_F$ the output~$p_S(\ga) =
      (\alpha,(a_\gp)_{\gp\in S})$ satisfies~$\ga = (\alpha)\prod_{\gp\in S}\gp^{a_\gp}$;
    \item for each modulus $\gm$, a lattice $\Lambda_\gm$ of rank $r(\gm)$
        and a logarithm oracle $\log_\gm\colon \Z_F\to\Z^{r(\gm)}$
        inducing an isomorphism~$(\Z_F/\gm)^\times \cong
        \Z^{r(\gm)}/\Lambda_\gm$.
\end{itemize}
\end{defi}

Note that these oracles are available in Pari/GP, using the algorithms described
in~\cite{Buchmann1990},\cite{CohenFirst}, \cite[Section 4.2]{CohenAdvanced}
and~\cite{DiscreteLogs}.

Using the notations introduced in Definition~\ref{def:mapL} and Proposition~\ref{prop:groupmodnorm},
our algorithms are the following.

\begin{algo}\label{algo:log}~
\begin{itemize}
    \item Input: a strongly computable number field $F$, a modulus~$\gm$ and an ideal $\ga\in I_F$.
    \item Output: a vector $z$ in $\R^{\ell+n}$ such that $\cL(\ga)\equiv z\bmod \Lambda$.
\end{itemize}
\begin{enumerate}
    \item Let $(\alpha,(a_\gp)_\gp)=p_S(\ga)$.
    \item Let $u\in \Z_F$ such that $\alpha\pi_\gp^{-v_\gp(\alpha)}\equiv
      u\bmod\gp^{v_\gp(\gm_f)}$ for all $\gp\mid\gm_f$.
    \item Return $z=((a_\gp)_{\gp\in S},-\log_\gm(u),-\log_\infty(\alpha))$.
\end{enumerate}
\end{algo}

\begin{algo}\label{algo:group}~
\begin{itemize}
    \item Input: a strongly computable number field $F$ and a modulus $\gm$.
    \item Output: a matrix $B$ whose rows generate $\widehat{C}^1_\gm$ in $\R^{\ell+n}$.
\end{itemize}
\begin{enumerate}
    \item Let $A$ be a matrix whose columns form a basis
        of $\cL_S(\Z_{F,S}^\times)+\Lambda_\gm+\Z^{r_2}+\Z v_0$ in $\R^{\ell+n}$.
    \item Let $B = A^{-1}$: the rows of~$B$ form the basis dual to the columns of~$A$.
    \item Delete from~$B$ the row corresponding to the linear form dual to~$v_0$.
    \item Replace the rows of~$B$ by their orthogonal projections onto $(\R v_0)^\perp$.
    \item Return the $(\ell+n-1)\times(\ell+n)$ matrix~$B$.
\end{enumerate}
\end{algo}

\begin{rmk}
These algorithms output numerical approximations in $\R^{\ell+n}$: their validity
to any prescribed accuracy can be certified as follows.
In both cases, the numerical approximations come from log embeddings of number
field elements, which can be obtained to arbitrary accuracy in polynomial time.
All subsequent numerical operations come from linear algebra and can be implemented using certified
numerical algorithms \cite{Johansson2017arb} with automatic precision increase until a target precision is reached.
Our package implements this strategy except that we rely on Pari/GP's arithmetic which is not
certified.
\end{rmk}

\begin{thm}
    Algorithm~\ref{algo:group} and Algorithm~\ref{algo:log} are correct.
    They are polynomial time, meaning a polynomial number of calls to the
    oracles with polynomial size input and a polynomial number of other
    operations.
\end{thm}
\begin{proof}
    Algorithm~\ref{algo:group} is correct by Proposition~\ref{prop:groupmodnorm}.
    
    We verify that the value $z$ computed in Algorithm~\ref{algo:log} equals
    $\cL(\ga)\bmod\Lambda$: let $x=(\pi_\gp^{v_\gp(\ga)})$ be an idèle defining $\ga$,
    we have $\cL(\ga)=\cL(x)=\cL_S(x\alpha^{-1})$ by definition of $\cL$.
    Now we have
    $v_\gp(x\alpha^{-1})=a_\gp$ for $\gp \in S$ by definition of $p_S$,
    and $x\alpha^{-1}\equiv u^{-1}\bmod \gm$ by definition of $u$.
    At infinite places $(x\alpha^{-1})_\sigma=\alpha^{-1}_\sigma$.
    Hence $\cL_S(x\alpha^{-1})\equiv z\bmod \Lambda$, and Algorithm~\ref{algo:log} is
    correct.
    All operations not provided by the oracles can clearly be performed in
    polynomial time.
\end{proof}

\section{The subgroup of algebraic characters}\label{sec:algebraic}

Among Hecke quasi-characters, we would like to exhibit the subgroup of algebraic
Hecke characters.
By Lemma~\ref{lem:infinitytype}, it is equivalent to compute the subgroup of almost-algebraic
characters inside the group of Hecke characters.
More precisely, let~$H_0^\bot\subset\R^{\ell+n}$ be the subgroup of characters defined by
$%\[
  H_0^\perp = \set{\varphi_\sigma = 0\text{ for all }\sigma},
$ %\]
then
\[
  \begin{aligned}
      \Cmalg &= \widehat{C}_\gm\cap \Calg \\
             &\cong \Lambda^\perp\cap H_0^\perp/\Z^\ell
             =\set{\lambda\in\Lambda^\perp \mid \lambda(h)=1 \text{ for all }h\in H_0}/\Z^\ell.
  \end{aligned}
\]

However, we do not want to solve
the equation~$\varphi_\sigma=0$ since the components~$\varphi_\sigma$
on~$\Lambda^\perp$ are only known approximately.
We are therefore going to use the known structure of algebraic
characters.

Recall that a number field~$K$ is \emph{CM} if it is a totally complex quadratic
extension of a totally real field, denoted~$K^+$.
In this case, the automorphism corresponding
to this quadratic extension induces complex conjugation on every complex
embedding of~$K$, and we therefore denote it by~$x\mapsto \bar{x}$.

A classical theorem of Weil and Artin states the
following~\cite{Weil1956,Patrikis2019}:
\begin{itemize}
  \item If~$F$ does not admit a CM subfield, then every algebraic Hecke character is a
    finite order character times an integral power of the norm.
  \item If~$F$ admits a CM subfield, then it admits a maximal CM subfield~$K$.
    The type of every algebraic Hecke character of~$F$ is the lift of the type of an
    algebraic character of~$K$. Equivalently, every almost-algebraic Hecke
    character of~$F$, up to a finite order character, factors through the
    norm~$\Nm_{F/K}$ to~$K$.
\end{itemize}

%Below we
%describe an algorithm to compute the subgroup of almost-algebraic characters
%whose parameters at infinity factor through the norm to a given CM subfield~$K$.

\subsection{Determining the subgroup of algebraic characters from the maximal CM subfield}

In this section, we assume that~$F$ contains a CM subfield. In particular, $F$ is
totally complex.

Let~$G = \R^{\ell+n}$ be equipped with its standard inner product and
$\Lambda_0=\Lambda+\Z v_0=\cL_S(\Z_{F,S}^\times) +\Lambda_{\gm} + \Z^{r_2}+\Z v_0$,
so that~$\widehat{C}^1_\gm\times\|\cdot\|^{i\Z} \cong \Lambda_0^\bot/\Z^\ell$,
with $\Lambda_0^\bot=\Lambda_0^\vee$ in $G$.

Our strategy is to capture the algebraic characters in a
smaller subspace $H^\perp\subset G$ by using the additional known constraints on
almost-algebraic characters, in order to apply the following lemma.
\begin{lem}
  \label{lem:subgroupexact}
  Let~$G$ be a finite dimensional $\R$-vector space, let~$H\subset G$ be
  an~$\R$-vector subspace and let~$\Lambda_0\subset G$ be a lattice
  such that~$H\cap \Lambda_0$ has full rank in~$H$. Then
    \[ \Lambda_0^\bot\cap H^\bot=\set{\lambda\in \Lambda_0^\bot, \lambda\cdot h=0
    \text{ for all }h\in H\cap \Lambda_0}. \]
  %If in addition~$\Lambda$ has full rank in~$G$, then~$\Lambda^\bot\cap H^\bot$
  %is a full rank lattice in~$H^\perp$.
\end{lem}
\begin{proof}
    We use the fact that~$H$ is an~$\R$-subspace generated by~$H\cap \Lambda_0$ to write
    \[
        \begin{aligned}
        \Lambda_0^\bot\cap H^\bot &= \set{x\in \Lambda_0^\bot, x\cdot h\in\Z \text{ for all }h\in H}
                       \\&= \set{x\in \Lambda_0^\bot, x\cdot h = 0 \text{ for all }h\in H}
                       \\&= \set{x\in \Lambda_0^\bot, x\cdot h = 0 \text{ for all
                       }h\in H\cap \Lambda_0},
        \end{aligned}
    \]
    proving the claim.
    %proving the first claim.
    %Now suppose that~$\Lambda$ has full rank in~$G$.
    %First note that~$\Lambda^\perp\cap H^\perp$ is a lattice in~$H^\perp$.
    %On the one hand we have
    %\[
    %  \widehat{G}/H^\perp \cong \widehat{H},
    %\]
    %so~$\dim_\R G = \dim_\R \widehat{G} = \dim_\R H + \dim_\R H^{\perp}$.
    %On the other hand we have
    %\[
    %  \frac{\Lambda^\perp}{\Lambda^\perp \cap H^\perp}
    %  = \frac{\Lambda^\perp}{(\Lambda + H)^\perp}
    %  \cong \frac{\widehat{G/\Lambda}}{(H/H\cap\Lambda)^\perp}
    %  \cong \widehat{\left(\frac{H}{H\cap\Lambda}\right)}.
    %\]
    %Since~$\Lambda$ has full rank in~$G$ we have~$\dim_\R(G) = \rank_\Z(\Lambda)
    %= \rank_\Z(\Lambda^\perp)$, and since~$H\cap\Lambda$ has full rank in~$H$ we
    %have~$\dim_\R H = \rank_\Z(H\cap\Lambda) =
    %\rank_\Z(\widehat{\frac{H}{H\cap\Lambda}})$.
    %We get~$\dim_\R G = \rank_\Z(\Lambda^\perp \cap H^\perp) + \dim_\R H$,
    %so we obtain~$\rank_\Z(\Lambda^\perp \cap H^\perp) = \dim_\R(H^\perp)$,
    %which is exactly the full rank claim.
\end{proof}

\begin{rmk}
  The point of Lemma~\ref{lem:subgroupexact} is that since the inner products between elements
  of~$\Lambda_0$ and~$\Lambda_0^\perp$ are in~$\Z$, the given expression
  for~$\Lambda_0^\perp\cap H^\perp$ can be computed exactly as a subgroup
  of~$\Lambda_0^\perp$ by linear algebra over~$\Z$.
\end{rmk}

\begin{ex}
    When~$H_0^\perp =\set{\varphi_\sigma=0}$ as above, we have~$H_0 = \R^{r_2}$.
  Then $H_0\cap \Lambda_0$ is~$\cL_S(\Z_{K^+}^\times) + \Z v_0$, which has rank~$r_1(K^+)
  = r_2(K)$. This has full rank in~$H_0$ if and only if~$K=F$.
\end{ex}

This example shows that using~$H_0$ is sufficient when~$F$ itself is CM.
In the general case, we proceed as follows.

\begin{prop}\label{prop:algchar}
Let~$K$ be the maximal CM subfield of~$F$, let
\[
  H^\perp = \{\varphi_\sigma=0 \text{ for all }\sigma, \text{ and }
  (k_\sigma)_\sigma \text{ factors through }K\}
\]
    and~$\Lambda_0=\cL_S(\Z_{F,S}^\times) +\Lambda_{\gm} + \Z^{r_2} + \Z v_0$.
Then
\[
    \Cmalg = (\Lambda_0^\bot\cap H^\bot)/\Z^\ell.
\]
where~$\Lambda_0\cap H$ has full rank in~$H$. More precisely, the group~$U\subset
  \Lambda_0$
  generated by~$v_0$, the kernel~$\ker(\Nm_{F/K} \colon \Z^{r_2} \to \Z^{r_2(K)})$
  and~$\cL_S(u)$ for all~$u\in\ker (\Z_F^\times \to (\Z_F/\gm)^\times)$ 
  such that~$\Nm_{F/K}(u)\in K^+$, is contained in~$H$ and has full rank.
\end{prop}
\begin{proof}
  Almost-algebraic characters are contained in~$H^\perp$ since their infinity types factor
  through $\Nm_{F/K}$, and
  we have $H = \R^{r_2} \times \ker(\Nm_{F/K} : \R^{r_2} \to \R^{r_2(K)})$.
  The group~$U$ described in the Proposition is clearly contained
  in~$H\cap\Lambda_0$.
  The map~$\Nm_{F/K} \colon \Z^{r_2} \to \Z^{r_2(K)}$ is surjective since every
  complex place of~$K$ extends to a complex place of~$F$, so that its kernel has
  rank~$r_2-r_2(K)$. Finally, the units described form a finite index subgroup
  of~$\Z_F^\times$, so the group~$U$ has full rank in~$H$.
  %There exists a finite index subgroup of~$\Lambda$ having full rank
  %in~$\R^{2r_2}$, and image of full rank in the second factor~$\R^{r_2}$,
  %so~$H\cap \Lambda$ has full rank.
\end{proof}

% I think this is not useful in the current description of the algorithms.
%In order to have a completely rational description of $\Cmalg$ our
%algorithm will take advantage of the following rationality result.
%\begin{lem}
%  \label{lem:int-narg}
%  Let~$K$ be a CM field and let~$\rootsofone{K}$ be the number of roots of unity in~$K$. Then for
%  every~$u\in\Z_K^\times$ and~$\sigma\colon K\hookrightarrow\C$, we have
%  \[
%    \frac{\arg \sigma(u)}{2\pi} \in \frac{1}{2\rootsofone{K}}\Z.
%  \]
%\end{lem}
%\begin{proof}
%  Let~$z = u/\bar{u}\in\Z_K^\times$. Then for every complex embedding~$\sigma$,
%  we have~$|\sigma(z)|=1$. So~$z$ is a root of unity: $z^{\rootsofone{K}}=1$.
%  We obtain
%  \[
%    2\arg(\sigma(u)) = \arg(z) \in \frac{2\pi}{\rootsofone{K}}\Z,
%  \]
%  hence the result.
%\end{proof}

%For us: $L_1\subset L_2$ lattices with~$L_1$ saturated (so that~$L_2/L_1$ is a lattice).
%In~$L_2^\vee$, we have~$L_1^\perp = (L_2/L_1)^\vee$.

\subsection{The maximal CM subfield}\label{sec:cmsubfield}

In this section we reformulate the problem of determining the maximal CM
subfield in a way that is suitable for an efficient algorithm. Indeed
enumerating all subfields, regardless of the algorithm used, could not lead to a
polynomial time algorithm since the number of subfields is not polynomially
bounded, as the example of multi-quadratic fields shows. One may consider a pure Galois-theoretic approach, but it is currently
not known whether one can compute in polynomial time, given a number field~$F$,
the Galois group of the Galois closure of~$F$ (see \cite{ComplexityGalois2003,SolvabilityRadicals1985,ElsenhansKluners2019}).
Our method relies on the following Lemma.

\begin{lem}\label{lem:cmsubfield}
  Let~$F$ be a number field. For~$\eps\in\{\pm\}$, let
  \[
    F^\eps = \{x\in F\mid \sigma(x) = \eps\bar\sigma(x) \text{ for all }\sigma\in\Hom(F,\C)\}.
  \]
  The following are equivalent:
  \begin{enumerate}[(i)]
    \item $F$ admits a CM subfield;
    \item $F^-\neq 0$;
    \item $\dim_\Q F^+ = \dim_\Q F^-$.
  \end{enumerate}
  If those conditions are satisfied, then the largest CM subfield of~$F$ is~$F^+
  + F^-$; it also equals~$\Q(a)$ for every~$a\in F^-$ having minimal
  polynomial of degree~$2\dim_\Q F^-$, and such an element exists.
\end{lem}
\begin{proof}
  First note that~$F^+$ is the largest totally real subfield of~$F$.
  It is clear that~(i) implies~(ii).
  Since~$\dim_\Q F^+\ge 1$, (iii) implies~(ii).
  Let~$a,b\in F^-$ be nonzero; then~$a/b\in F^+$ and therefore~$F^-$ is a one-dimensional vector space over~$F^+$, so~(ii) implies~(iii).
  Let~$a\in F^-$ be nonzero; then~$a^2\in F^+$ is totally negative, so~$F^+(a) = F^+ + F^-$ is a CM subfield of~$F$, so that~(ii) implies~(i).
  If the conditions are satisfied, then the maximal CM subfield~$K$ of~$F$ is a quadratic extension of its totally real subfield~$F^+$ containing~$F^+ + F^-$,
  so there is equality as claimed. Let~$a\in F^-\subset K$ have minimal polynomial of
  degree~$2\dim_\Q F^- = [K:\Q]$; then it generates~$K$ over~$\Q$.
  For every subfield~$L\subset K$, if~$F^-\subset L$ then~$F^+\subset L$ by
  taking ratios, so~$K\subset L$ and therefore~$L = K$. The set of elements
  of~$F^-$ lying in a proper subfield of~$K$ is therefore a finite union of
  proper subspaces, and is therefore nonempty.
\end{proof}

It is therefore enough to compute~$F^-$. Proposition~\ref{prop:twoemb}
below gives a general algorithm to solve this type of problem.

\begin{prop}\label{prop:twoemb}
  Let~$F$ be a number field.
  Let~$\Omega$ be a field of characteristic~$0$, let~$R\subset \Hom(F,\Omega)^2$
  be a subset and let~$(\lambda_r)_{r\in R} \in \Q^R$ be a family of rational numbers.
  Define
  \[
    F_{R,\lambda} = \{x\in F \mid \sigma_1(x) = \lambda_r \sigma_2(x) \text{ for all } r=(\sigma_1,\sigma_2)\in R\}.
  \]
  Write~$F\otimes_\Q F \cong \prod_{i=1}^k L_i$ where each~$L_i$ is a field.
  Let~$p_i \colon F\otimes_\Q F \to L_i$ be the projection onto~$L_i$.
  %Moreover, every element of~$\Hom(L_i,\Omega)$ is injective.
  For each~$r\in R$, let~$i(r)\in\{1,\dots,k\}$ be the index such that~$r$ corresponds to an element of~$\Hom(L_i,\Omega)$ under the natural bijection
  \[
    \Hom(F,\Omega)^2 \cong \Hom(F\otimes_\Q F, \Omega) \cong \bigsqcup_{i=1}^k \Hom(L_i,\Omega),
  \]
  where the last union is disjoint.
  Let~$f : F \to \bigoplus_{r\in R}L_{i(r)}$ be the~$\Q$-linear map defined by
  \[
    f(x)_r = p_{i(r)}\bigl(x\otimes 1 - \lambda_r(1\otimes x)\bigr) \text{ for all }r\in R.
  \]
  Then~$F_{R,\lambda} = \ker f$.
\end{prop}
\begin{proof}
  Let~$i\in\{1,\dots,k\}$ and~$\varphi \in \Hom(L_i,\Omega)$ correspond to~$(\sigma_1,\sigma_2)\in\Hom(F,\Omega)^2$.
  Then, for all~$x\in F$, we have~$\sigma_1(x) = \varphi(p_i(x\otimes 1))$ and~$\sigma_2(x) = \varphi(p_i(1\otimes x))$.
  Noting that~$\varphi$ is injective since~$L_i$ is a field, we obtain for every~$\lambda\in\Q$ the equivalence
  \[
    \sigma_1(x) = \lambda\sigma_2(x)
    \Leftrightarrow
    \varphi\left(p_i\bigl(x\otimes 1-\lambda (1\otimes x)\bigr)\right)=0
    \Leftrightarrow
    p_i\bigl(x\otimes 1 - \lambda (1\otimes x)\bigr)=0.
  \]
  This proves the claim.
\end{proof}

The advantage of rewriting the equations this way is that instead of having conditions in~$\Omega$
(which might be a field in which we cannot compute exactly such as~$\Omega = \C$
or~$\Omega = \overline\Q_p$),
the conditions take place in the number fields~$L_i$ and~$f$ is a linear map between
finite-dimensional~$\Q$-vector spaces.

\begin{rmks}~
  \begin{itemize}
    \item There are obvious generalizations to conditions expressed with more
      than two embeddings, but they become more and more expensive as the number
      of embeddings increases; eventually one may have to compute the full
      Galois closure of~$F$.
    \item The application to the maximal CM subfield can be generalized to other
      natural conditions, such as the maximal real subfield, the maximal
      subfield fixed by some ramification group, or the maximal subfield in
      which the residue degree of a certain prime divides a given integer.
    \item When~$\lambda_r=1$ for all~$r\in R$, Proposition~\ref{prop:twoemb}
      expresses the subfields of interest as intersections of \emph{principal
      subfields} in the terminology of van Hoeij, Kl\"uners and
      Novocin~\cite{subfields}.
  \end{itemize}
\end{rmks}

\subsection{Algorithms}

Section~\ref{sec:cmsubfield} leads to the following algorithm to compute the
maximal CM subfield.

\begin{algo}\label{algo:cmsubfield}~
  \begin{itemize}
    \item Input: an irreducible monic~$P\in\Q[X]$ representing~$F=\Q[X]/(P(X))$.
    \item Output: an element~$a\in F$ such that~$\Q(a)$ is the maximal CM
      subfield of~$F$, or~$\bot$ if~$F$ does not contain a CM subfield.
  \end{itemize}
  \begin{enumerate}
    \item Let~$P(Y) \equiv \prod_{i} Q_i(X,Y) \bmod P(X)$ be the irreducible factorization of~$P$ over~$F$.
    \item Let~$J$ be the set of indices~$i$ such that there exists a complex
      root~$\alpha$ of~$P$ such that~$Q_i(\alpha,\bar\alpha)=0$.
    \item Let~$V\subset F$ be the~$\Q$-subspace of~$a(X)\bmod P(X)$ such that for all~$i\in J$,
        $a(X)+a(Y)\equiv0\bmod (P(X),Q_i(X,Y))$.
    \item If~$V=0$, return~$\bot$.
    \item Let~$a\in V$ be such that the minimal polynomial of~$a$ has
      degree~$2\dim_\Q V$. Return~$a$.
  \end{enumerate}
\end{algo}

\begin{thm}\label{thm:algocmsubfield}
  Algorithm~\ref{algo:cmsubfield} is a deterministic polynomial-time algorithm
  that, given a number field~$F$, computes the maximal CM subfield of~$F$.
\end{thm}
\begin{proof}
  Algorithm~\ref{algo:cmsubfield} is correct by Lemma~\ref{lem:cmsubfield} and
  Proposition~\ref{prop:twoemb} since~$F\otimes_\Q F \cong \Q[X,Y]/(P(X),P(Y))$.
  It runs in polynomial time because factorization of polynomials over number
  fields can be performed in polynomial time~\cite{nffactor}. %also cite lll?
\end{proof}

We obtain the following algorithm to compute the group of almost-algebraic
characters.

\begin{algo}\label{algo:algchar}~
  \begin{itemize}
    \item Input: a strongly computable number field~$F$ and a modulus~$\gm$.
    \item Output: the group of almost-algebraic characters of modulus~$\gm$.
  \end{itemize}
  \begin{enumerate}
    \item Let~$K$ be the maximal CM subfield of~$F$, as computed by
      Algorithm~\ref{algo:cmsubfield}.
    \item If~$K = \bot$, return the group of finite order characters.
    \item Let~$A,B$ be the matrices computed by Algorithm~\ref{algo:group} with
      input~$(F,\gm)$.
    \item Let~$U$ be the subgroup described in Proposition~\ref{prop:algchar}.
    \item Let~$C$ be the subgroup of the row span of~$B$, consisting
      of elements~$c$ such that~$u\cdot c = 0$ for all~$u\in U$.
    \item Output~$C$.
  \end{enumerate}
\end{algo}

\begin{thm}
    Algorithm~\ref{algo:algchar} is correct.
    It is polynomial time, meaning a polynomial number of calls to the
    oracles with polynomial size input and a polynomial number of other
    operations.
\end{thm}
\begin{proof}
  If~$F$ does not contain a CM subfield, then almost-algebraic characters are
  exactly finite order characters by the Artin--Weil theorem.
  The group~$U$ can be computed by linear algebra using the oracles.
  The group~$C$ can be computed by linear algebra over~$\Q$ since all the inner
  products that occur are in~$\Z$. The group~$C$ is the correct output by the
  Artin--Weil theorem and Lemma~\ref{lem:subgroupexact} in combination with
  Proposition~\ref{prop:algchar}.
  All operations not provided by the oracles or Theorem~\ref{thm:algocmsubfield}
  can clearly be performed in polynomial time.
\end{proof}

%\subsection{Field of definition and algebraic evaluation}
% Won't do (no time).

\section{Examples}
\label{sec:examples}

We illustrate the interface of our Pari/GP package \cite{parigp:gchar}
with a list of examples of mathematical interest.

\subsection{Pari/GP interface}

The \kbd{gcharinit(F,m)} function initializes a group structure~\kbd{gc}
for a number field~$F$ and a modulus~$\gm$.
The character group structure
$\widehat{C}_\gm \cong \prod_{i=1}^k\Z/c_i\Z\times \Z^{n-1}\times\R$
is obtained via the vector \kbd{gc.cyc} $= [c_1,\dots c_k, 0, \dots, 0, 0.]$.

As an example,
\begin{code}
> gc = gcharinit(x^2+23,3);
> gc.cyc
[6, 0, 0.E-57]
\end{code}
expresses the group of Hecke quasi-characters of modulus~$\gm=(3)$
over~$F=\Q(\sqrt{-23})$ (see also Equation~(\ref{eq:normsplitting}))
\[
    \Hom_{\rm cont}(C_3,\C^\times) = \chi_3^{\Z/6\Z}\times\chi_{CM}^\Z\times\|\cdot\|^\C,
\]
where~$\chi_3$ is a character of the ray class group~$\Cl_F(3)$ and
$\chi_{CM}$ is an infinite order almost-algebraic character.% of CM type.

Characters are described as columns of coordinates in this basis.
\begin{code}
> gchareval(gc,[1,0,0]~,idealprimedec(gc.nf,3)[1])
-0.5000 - 0.8660*I \\ the prime above 3 is not principal
> gcharconductor(gc,[2,0,0]~)
[1, []] \\ a class group character
\end{code}

The maps~$\cL$ and~$\cL^*$ are accessible as \kbd{gcharlog} and
\kbd{gcharduallog}, except that these functions have an extra component
corresponding to the norm. For example the character~$\chi_{CM}$ has
the following parameters in $\bigl((\R/\Z)^3\times\R\times\Z\bigr)\times\C$,
where:
\begin{itemize}
  \item the set~$S$ is~$\bigl\{(2,\frac{\sqrt{-23}-1}{2})\bigr\}$;
  \item the map~$\log_\gm\colon (\Z_F/\gm)^\times \to \Z^2/2\Z^2$ is characterized
    by~$\log_\gm(2) = (1,1)$ and~$\log_\gm(\sqrt{-23}) = (1,0)$.
\end{itemize}

\begin{code}
> gcharduallog(gc,[0,1,0]~)
[0.11298866677205092301511538301498585720, 0, 1/2, 0, 1, 0]
\end{code}
%The fact that this infinite order character has a first
%non trivial component on $\Cl_F$ illustrates the general
%fact that the exact sequence~(\ref{eq:usualsequence}) is
%not a direct product for the canonical embeddings.

For closer scrutiny we retrieve the local quasi-characters
of $\chi = \chi_{CM}\|\cdot\|$. In
particular for a prime $\gp_3$ dividing the conductor $\gm=3$
we obtain a character of the \kbd{idealstar} structure $(\Z_F/\gp_3)^\times$ in addition
to a value~$\theta\in\C$ such that~$\chi(\gp_3) = \exp(2\pi i \theta)$.
\begin{code}
> gcharlocal(gc,[0,1,1]~,1) \\ complex place
[1, -I] \\ k = 1, phi = -I
> gcharlocal(gc,[0,1,1]~,idealprimedec(gc.nf,3)[2],&grp)
[1, 0.1042940216...+ 0.1748495762...*I] \\ [grp char, theta]
> grp.cyc
[2] \\structure of (ZF/p3)^*
\end{code}

%so that for any ideal $\ga = \alpha \gp^i$ we have
%$\chi_{CM}(\ga) = i(\frac{\alpha}{\abs{\alpha}})^2$
The interface gives a basis of the subgroup of algebraic characters.
We can work with these characters via their type.
\begin{code}
> Vec(gcharalgebraic(gc))
[[1, 0, 0]~, [0, 1, -1/2]~, [0, 0, -1]~]
> gcharisalgebraic(gc,[2,-3,5/2]~,&t); t
[[-1, -4]] \\ type (-1,-4)
> gcharalgebraic(gc,[[-1,2]])
[[0,3,-1/2]~] \\ an algebraic character of type (-1,2)
\end{code}

The $L$-function machinery is readily accessible.
\begin{code}
> lfunzeros([gc,[1,3,0]~],5)
[2.34520501265099..., 3.90705697239550...]
> lfunan([gc,[0,3,-3/2]~],8)
[1,4.795...*I,2+4.795...*I,-15,0,-23+9.591...*I,0,-33.570...*I]
> [ algdep(an,2) | an <- % ] \\ check algebraicity
[x-1, x^2+23, x^2-4*x+27, x+15, x, x^2+46*x+621, x, x^2+1127]
\end{code}
%[1, 4.795...*I, 2+4.795...*I, -15, 0, -23+9.591...*I, 0, -33.570...*I]

\ifarxiv
\subsection{Small degree examples}

We describe explicitly the form
taken by infinite order Hecke characters
and our choice of basis for low degree fields.

We denote
$z=(z_1,\dots, z_{r_1+r_2})$ the elements
of $F_\R\simeq\R^{r_1}\times\C^{r_2}$.
The characters of $(F_\R^\times)^\cc$ are of the form
\[
    \chi_\infty(z) = \prod_{j=1}^{r_1+r_2}\abs{z_j}^{i n_j\varphi_j}\prod_{j=r_1+1}^{r_1+r_2}\left(\frac{z_j}{\abs{z_j}}\right)^{k_j}
\]
and conversely, such a character $\chi_\infty$ can be extended to
a global Hecke character if
it is trivial on a finite index subgroup of~$\Z_F^\times$.% up to the twist by a finite order character.

Working modulo the norm, we therefore consider the characters of
\[
  G^1_\infty=(F_\R^\times)^\cc/(\Z_F^\times\cdot \R_{>0})
\]
where~$\R_{>0}$ is embedded diagonally.
\newcommand{\hGinfty}{\widehat{G^1_\infty}}
The group~$\hGinfty$ is free of rank~$n-1$ and is a full rank lattice in
the~$\Q$-vector space of all possible parameters at infinity.

When~$F$ has class number one and totally positive fundamental units,
$\hGinfty$ is precisely the lattice of infinite order characters of modulus~$\gm=1$.

\begin{ex}
For $F=\Q$, infinite order characters are powers of the norm,
    and finite order characters are Dirichlet characters.
\end{ex}

\begin{ex}[real quadratic]
Let $F=\Q(\sqrt{D})$ be real quadratic with fundamental unit
$\eta_1>1$ and regulator $R_F=\log(\eta_1)$.
    Then $\hGinfty$ is generated by
\[
  \chi(z)=\abs{\frac{z_1}{z_2}}^{i\frac{\pi}{R_F}}.
\]
\end{ex}

\begin{ex}[imaginary quadratic]
Let $F=\Q(\sqrt{D})$ be imaginary quadratic with torsion units
of order~$m$. Then $\hGinfty$ is generated by
\[
    \chi(z)=\Bigl(\frac{z_1}{\abs{z_1}}\Bigr)^{m}.
\]
\end{ex}

\begin{ex}[complex cubic]
Let $F$ be complex cubic,
%choose a fundamental unit whose real embedding
%is $\eta>1$, and write $\eta'=\frac{e^{2i\pi\alpha}}{\sqrt{\eta}}$
%its complex embedding. The regulator is $R=\log(\eta)$.
%
%We obtain a basis $\chi_1=(\frac{2}{3R},-\frac{2}{3R},0)$
%and $\chi_2=(-\frac{2\alpha}{3R},\frac{2\alpha}{3R},1)$.
and consider a fundamental unit whose complex embeddings are
$e^{-\frac{R_F}2\pm 2i\pi\alpha}$, where $R_F>0$ is
the regulator and $\alpha\in \R/\Z$ is an angle.

Then $\hGinfty$ is generated by
% [r,-r,a;0,0,1/2;1,2,0]~^-1
\[
    %\chi_1(z) = \abs{\frac{z_1}{z_2^2}}^{2i\pi\frac{2}{3 R_F}}
    \chi_1(z) = \abs{\frac{z_1}{z_2}}^{2i\pi\frac{2}{3 R_F}}
\]
and
\[
    %\chi_2(z) = \abs{\frac{z_1}{z_2^2}}^{-2i\pi\frac{4\alpha}{3 R_F}}
    \chi_2(z) = \abs{\frac{z_1}{z_2}}^{-2i\pi\frac{4\alpha}{3 R_F}}
    \left(\frac{z_2}{\abs{z_2}}\right)^2.
\]
\end{ex}

\begin{ex}[real cubic]
Let $F$ be real cubic, and $(\pm e^{\alpha_i})_i,(\pm e^{\beta_i})_i\in F_\R$
the embeddings of two fundamental units,
so that the regulator is~$R_F=\abs{\alpha_1\beta_2-\alpha_2\beta_1}$.
Then $\hGinfty$ is generated by
% [a1,a2,-a1-a2;b1,b2,-b1-b2;1,1,1]~^-1
\[
    \chi_1(z) =
    \abs{z_1}^{2i\pi\frac{\alpha_1+2\alpha_2}{3R_F}}
    \abs{z_2}^{2i\pi\frac{-2\alpha_1-\alpha_2}{3R_F}}
    \abs{z_3}^{2i\pi\frac{\alpha_1-\alpha_2}{3R_F}}
\]
and
    \[
    \chi_2(z) =
    \abs{z_1}^{2i\pi\frac{\beta_1+2\beta_2}{3R_F}}
    \abs{z_2}^{2i\pi\frac{-2\beta_1-\beta_2}{3R_F}}
    \abs{z_3}^{2i\pi\frac{\beta_1-\beta_2}{3R_F}}.
\]
\end{ex}

%Let $F$ be the degree 4 field of discriminant $117$.
%
%However this field has CM by $\Q(\zeta_3)$, hence the algebraic
%Hecke characters live in the lattice
\fi

\subsection{Modular forms}

%automorphic principle states that (proved for cyclic extensions $F/K$)

By automorphic induction, Hecke characters of an extension $F/K$
are expected to induce automorphic representations of $\GL_{[F:K]}$ over $K$.
This is known in a number of cases. Here we provide some explicit examples for
quadratic fields, where converse theorems prove the existence of a global
automorphic form.

\subsubsection{Classical forms over $\GL_2$}

Let $F=\Q(\sqrt{-D})$ be an imaginary quadratic field of discriminant $-D<0$ and $k>0$. To
an algebraic character $\chi$ of type $(k,0)$
and conductor $\gm$ we associate the $q$-series
\[
    f_\chi(z) = \sum_{(\ga,\gm)=1}\chi(\ga)q^{\Nm(\ga)}, q=e^{2i\pi z}, \im(z)>0
\]
where the sum runs over integral ideals $\ga$ coprime
to $\gm$.
\begin{thm}[Hecke\cite{HeckeWork}, Weil\cite{WeilConverse}, Shimura\cite{ShimuraEllCM,ShimuraConductorEllCM}]
    Let $\chi$ be an algebraic character of type $(k,0)$
    and conductor $\gm$ over $F=\Q(\sqrt{-D})$, then
    \[
        f_\chi\in S_{k+1}(\Gamma_0(N,\psi_F\psi_\chi))
    \]
    is a newform of weight $k+1$, level $N=D\Nm_{F/\Q}(\gm)$
    and character $\psi_F\psi_\chi$ where
    $\psi_F = \bigl(\frac{-D}\cdot\bigr)$ is the quadratic character of $F$
    and $\psi_\chi(a) = a^{-k}\chi((a))$ is the Dirichlet character of modulus
    $\Nm_{F/\Q}(\gm)$ attached to $\chi$.
\end{thm}

In the other direction, Ribet proved that all CM newforms come from algebraic Hecke
characters \cite[Theorem 4.5]{RibetCM}.
%%ref from https://arxiv.org/pdf/math/0511228.pdf
%\begin{thm}[Ribet, Prop. (4.4), Thm. (4.5)]
%A newform has CM by a quadratic field $F$ if and only if it comes from a Hecke character
%of $F$. In particular, the field K is imaginary and unique.
%\end{thm}

\newcommand\mfref[1]{\lmfdbref{#1}{ModularForm/GL2/Q/holomorphic/#1}}
\newcommand\dirref[1]{\lmfdbref{#1}{Character/Dirichlet/#1}}
\begin{ex}
    Consider $F=\Q(\sqrt{-19})$ and $\gm=3$. Our implementation show that up to
    integral powers of the norm,
    the algebraic characters are of the form $\chi_3^i\chi_\infty^k$
    where $\chi_3$ has order~$4$ and generates $\widehat{\Cl(\gm)}$, and
    $\chi_\infty$ has type $(1,0)$.
    In Table~\ref{tab:mf} we list the first algebraic characters and the corresponding
    CM modular forms referenced in \cite{lmfdb:modularforms}.
    \begin{table}\centering
        \begin{tabular}[t]{llll}
            %\toprule
            $(i,k)$ & quasi-character & modular form & first zero \\
            \hline
            %\midrule
            $(1,0)$ & \texttt{[1,0,0]}     & \mfref{171.1.c.a.37.1}  & $2.55662379\dots$ \\
            $(2,0)$ & \texttt{[2,0,0]}     & Dirichlet \dirref{57.56} & $2.40313422\dots$ \\
            $(3,0)$ & \texttt{[3,0,0]}     & \mfref{171.1.c.a.37.1}  & $2.55662379\dots$ \\
            $(0,1)$ & \texttt{[0,-1,-1/2]} & \mfref{171.2.d.a.170.3} & $1.19761556\dots$ \\
            $(1,1)$ & \texttt{[1,-1,-1/2]} & \mfref{171.2.d.a.170.1} & $3.03101717\dots$ \\
            $(2,1)$ & \texttt{[2,-1,-1/2]} & \mfref{171.2.d.a.170.2} & $2.19220898\dots$ \\
            $(3,1)$ & \texttt{[3,-1,-1/2]} & \mfref{171.2.d.a.170.4} & $0.57935987\dots$ \\
            $(0,2)$ & \texttt{[0,-2,-1]}   & \mfref{171.3.c.d.37.2}  & $1.76815328\dots$ \\
            $(1,2)$ & \texttt{[1,-2,-1]}   & \mfref{171.3.c.a.37.1}  & $1.84559250\dots$ \\
            $(2,2)$ & \texttt{[2,-2,-1]}   & \mfref{171.3.c.d.37.1}  & $1.54865425\dots$ \\
            $(3,2)$ & \texttt{[3,-2,-1]}   & \mfref{19.3.b.a.18.1}   & $3.78194741\dots$ \\
            $(0,3)$ & \texttt{[0,-3,-3/2]} & \mfref{171.4.d.a.170.4} & $1.59003776\dots$ \\
            $(1,3)$ & \texttt{[1,-3,-3/2]} & \mfref{171.4.d.a.170.3} & $1.36085197\dots$ \\
            $(2,3)$ & \texttt{[2,-3,-3/2]} & \mfref{171.4.d.a.170.1} & $0.08123213\dots$ \\
            $(3,3)$ & \texttt{[3,-3,-3/2]} & \mfref{171.4.d.a.170.2} & $0.70404412\dots$ \\
            %\bottomrule
        \end{tabular}
        \caption{Some modular forms with CM by $\Q(\sqrt{-19})$}
        \label{tab:mf}
    \end{table}
\end{ex}

\subsubsection{Maass waveforms}

Let $F=\Q(\sqrt{D})$ be a real quadratic field of discriminant $D$ and
fundamental unit $\eta_1>1$, and $\chi_m$ a Hecke character of
conductor $\gm=(\infty_1\infty_2)^\epsilon$ for $\epsilon\in\set{0,1}$
whose restriction to $F_\R^\times$ is
\[
    \chi_m(z)=\sgn(z_1z_2)^\epsilon\abs{\frac{z_1}{z_2}}^{ir_m},\, r_m = \frac{m\pi}{2\log(\eta_1)},
\]
where $\epsilon\equiv m\bmod 2$.

It corresponds to a CM Maass form
\cite[section 15.3.10]{CohenStromberg}.
\begin{prop}
Let $\cos^{(0)}(x)=\cos(x)$ and $\cos^{(-1)}(x)=\sin(x)$, and
$K_{ir}$ denote the modified Bessel function of the second kind of parameter $ir$.
The function
\begin{equation}
    f(x+iy) = \sqrt{y}\sum_{\ga} \chi_m(\ga) K_{ir_m}(2\pi \Nm(\ga)y)\cos^{(-\epsilon)}(2\pi \Nm(\ga)x)
\end{equation}
    is a cusp form of weight $0$ and character $\psi_F$ on $\Gamma_0(D)$ with Laplace
    eigenvalue $\lambda_m=\frac14+r_m^2$, % \nu_m = ir_m in Fredrik's notation
    where $\psi_F = \bigl(\frac{D}\cdot\bigr)$ is the quadratic character of $F$.
\end{prop}

\begin{ex}
    Let $F=\Q(\sqrt5)$, this field has trivial class group and fundamental unit
    $\eta=\frac{1+\sqrt5}2$. The character $\chi_m$ above is an actual
    Hecke character of modulus $\gm=(\infty_1\infty_2)^\epsilon$.
    Using the $L$-function facilities in Pari/GP we compute
    the first zero $0<\gamma_1$ such that $L(\chi_m,\tfrac12+i\gamma_1)=0$.
    Results are shown in Table~\ref{table:maass}.

   \begin{table}[h]\centering
       \begin{tabular}[t]{lll}
        %\toprule
        $m$ & $r_m=\frac{\pi m}{2\log(\eta_1)}$ & first zero \\
        %\midrule
        \hline
1   & $ 3.2642513026\dots$ & $7.4947673145\dots$ \\
2   & $ 6.5285026053\dots$ & $1.9926333454\dots$ \\
3   & $ 9.7927539079\dots$ & $1.3437292832\dots$ \\
4   & $13.0570052105\dots$ & $1.3684744255\dots$ \\
5   & $16.3212565132\dots$ & $0.9723034858\dots$ \\
6   & $19.5855078158\dots$ & $1.2974789657\dots$ \\
7   & $22.8497591185\dots$ & $0.7849215584\dots$ \\
8   & $26.1140104211\dots$ & $1.1328362023\dots$ \\
9   & $29.3782617237\dots$ & $0.8591419101\dots$ \\
10  & $32.6425130264\dots$ & $0.8952928125\dots$ \\
11  & $35.9067643290\dots$ & $0.7861064128\dots$ \\
12  & $39.1710156316\dots$ & $1.1315449163\dots$ \\
13  & $42.4352669343\dots$ & $0.5067080421\dots$ \\
14  & $45.6995182369\dots$ & $0.9758042566\dots$ \\
15  & $48.9637695395\dots$ & $0.8620736129\dots$ \\
%16 & 52.2280208422 & 0.7273833424 \\
%17 & 55.4922721448 & 0.6786040867 \\
%18 & 58.7565234475 & 0.9214119691 \\
%19 & 62.0207747501 & 0.5865227398 \\
%20 & 65.2850260527 & 0.9390964648 \\
%21 & 68.5492773554 & 0.4989814036 \\
%22 & 71.8135286580 & 0.7463975349 \\
%23 & 75.0777799606 & 2.4045155812 \\
%24 & 78.3420312633 & 0.7716034264 \\
%25 & 81.6062825659 & 0.3522523735 \\
%26 & 84.8705338685 & 0.8665737966 \\
%27 & 88.1347851712 & 0.7350082069 \\
%28 & 91.3990364738 & 0.6519048898 \\
%29 & 94.6632877765 & 0.6813532274 \\
%30 & 97.9275390791 & 0.7307166203 \\
%\bottomrule
    \end{tabular}
    \caption{First zero of Maass form $L$-functions of real quadratic field $\Q(\sqrt 5)$.}
    \label{table:maass}
    \end{table}
    Note that we obtain arbitrary large imaginary spectral parameters: this
    raises computational issues on the $L$-function side
    which are currently not addressed in Pari/GP.
    See \cite{BookerThen} for the case of degree 2 Maass forms.
 \end{ex}

\subsection{CM abelian varieties}% and motives}

% pour plus de courbes avec Jacobienne CM:
% https://arxiv.org/abs/1701.06489 et références dedans

% Fargues Motifs abéliens p20 & Cor7.9:
%
% If M is potentially CM motive over Q in the sens of Hodge cycles then
% it is modular in a weak sens : L(s,M) can be written as a product/quotient
% of $L$-functions of Hecke characters of finite extensions of Q
% (it is an easy adaptation of Brauer’s theorem to the Weil group that any
% representations of WQ is virtually a sum with Z-coefficients of inductions
% of a Hecke character)

In this section we give examples of CM abelian varieties and the
corresponding algebraic Hecke characters. We insist on proving equalities of
$L$-functions rather than observing a numerical coincidence, as this is possible
thanks to CM theory.
For the general terminology of CM theory, we refer to~\cite{LangCM,MilneCM}.
The following is a special case of~\cite[Chapter~4 Theorem~6.2]{LangCM}.

\begin{thm}[Shimura~\cite{ShimuraZetaCM}, Milne~\cite{MilneZetaCM}]\label{thm:cm}
  Let~$A/\Q$ be a simple abelian variety of dimension~$g$.
  Let~$K$ be a CM field of degree~$2g$ and~$\iota \colon K \to \End^0(A)$ an
  embedding, and let~$\Phi$ be the corresponding CM type on~$K$.
  Let~$F$ be the field of definition of~$\iota(K)$, and let~$\Phi^*$ be the dual
  type on~$F$.
  Then~$F/\Q$ is Galois; let~$G = \Gal(F/\Q)$.
  Let~$\pi$ be the injective morphism~$\pi \colon G \to \Aut(K)$ such
  that~$\iota(\lambda)^\sigma = \iota(\lambda^{\pi(\sigma)})$ for
  all~$\lambda\in K$ and~$\sigma\in G$.
  Then there exists an algebraic Hecke character~$\chi$ over~$F$ of
  type~$\Phi^*$ and valued in~$K$ such that
  \[
    L(A,s) = \prod_{\tau\in \Hom(K,\C)/\pi(G)} L(\chi^\tau, s).
  \]
\end{thm}

\begin{ex}
  Let~$A$ be the Jacobian of the genus~$2$ curve
  \lmfdbref{28561.a.371293.1}{Genus2Curve/Q/28561/a/371293/1}
  from the LMFDB~\cite{lmfdb:genus2}
  \[
    y^2 + x^3y = -2x^4 - 2x^3 + 2x^2 + 3x - 2.
  \]
  Let~$K = \Q[x]/(x^{4} - x^{3} + 2 x^{2} + 4 x + 3) = \Q(\alpha)$ be the unique degree~$4$
  subfield of~$\Q(\zeta_{13})$.
  %Since~$E/\Q$ is Galois, we have~$E^* = E$ regardless of~$\Phi$.
  The surface~$A$ is simple, has CM by~$K$, and all endomorphisms of~$A$ are
  defined over~$K$, as recorded in the LMFDB and proved by the
  algorithms of~\cite{CostaMascotGenus2,LombardoGenus2}.
  We therefore have~$F=K$ in the notation of Theorem~\ref{thm:cm}.
  %Moreover, the endomorphism ring of~$A$ over~$\Q$ is trivial, and therefore by
  %Falting's theorem, the representation~$\rho_\ell$ is absolutely irreducible
  %for all primes~$\ell$.
  Since~$K/\Q$ is Galois, $\pi(G)$ acts transitively on~$\Hom(K,\C)$.
  All CM types of~$K$ are in the same Galois orbit; let~$\Phi^* = \{\alpha
  \mapsto -0.65\ldots + 0.52\ldots i, \alpha \mapsto 1.15\ldots + 1.72\ldots i \}$.
  By Theorem~\ref{thm:cm}, there
  exists an algebraic Hecke character~$\chi$ of~$K$ of type~$\Phi^*$ such that
  \[
    L(A,s) = L(\chi,s).
  \]
  The conductor of~$A$ is~$28561 = 13^4$, and the discriminant of~$K$ is~$2197 =
  13^3$. Moreover, $K$ has a unique prime~$\gp$ above~$13$, so the conductor
  of~$\chi$ must be~$\gp$.

  Using our implementation we compute the group of characters of
  modulus~$\gp$. The subgroup of finite order characters has order~$3$, and
  there exists an algebraic character, unique up to multiplication by a
  finite order character, of type~$\Phi^*$. Among the three algebraic characters
  of this type, two have a non-real $L$-function coefficient~$a_3$, and
  therefore cannot be~$\chi$. So~$\chi$ is the remaining one,
  which is uniquely characterized by its type and the approximate value
  \[
    \chi(\gq) = -1.65138\ldots - 0.52241\ldots i
  \]
  where~$\gq = (3,\alpha)$ (label~\texttt{3.1} as defined in~\cite{labels}).
  The restriction of~$\chi$ to~$(\Z_K/\gp)^\times$ has order~$2$.
  %on le savait car nontrivial et valeurs dans K dont seules racines de l'unité +-1.
  The values of $\chi$ at some prime ideals are given in
  Table~\ref{table:genus2}.

  \begin{table}[h]
    \begin{tabular}[t]{c|c|c}
      prime~$\gr$ &  $\chi(\gr)\in \C$ & $\chi(\gr)\in K$ \\ \hline
      \texttt{3.1} & $-1.65138\ldots - 0.52241\ldots i$
        & $-\frac{1}{3}\alpha^3 - \frac{2}{3}\alpha - 2$ \\
      \texttt{3.2} & $0.15138\ldots - 1.72542\ldots i$
        & $-\frac{1}{3}\alpha^3 + \alpha^2 - \frac{5}{3}\alpha - 1$ \\
      \texttt{3.3} & $-1.65138\ldots + 0.52241\ldots i$
        & $\alpha-1$ \\
      \texttt{3.4} & $0.15138\ldots + 1.72542\ldots i$
        & $ \frac{2}{3}\alpha^3 - \alpha^2 + \frac{4}{3}\alpha + 1$\\
      \texttt{13.1} & $\pm 3.60555\ldots$
        & $\pm \sqrt{13} = \pm (\frac{2}{3}\alpha^3 - \frac{2}{3}\alpha + 3)$ \\
      \texttt{16.1} & $-4$ & $-4$ \\
      \texttt{29.1} & $-3.45416\ldots - 4.13143\ldots i$
        & $-\frac{5}{3}\alpha^3 + 3\alpha^2 - \frac{7}{3}\alpha - 5$\\
      \texttt{29.2} & $1.95416\ldots + 5.01809\ldots i$
        & $2\alpha^3 - 2\alpha^2 + 5\alpha + 5$ \\
      \texttt{29.3} & $-3.45416\ldots + 4.13143\ldots i$
        & $\frac{2}{3}\alpha^3 - 3\alpha^2 + \frac{10}{3}\alpha - 1$ \\
      \texttt{29.4} & $1.95416\ldots - 5.01809\ldots i$
        & $-\alpha^3 + 2\alpha^2 - 6\alpha - 2$
    \end{tabular}
    \caption{Values of the algebraic character~$\chi$ attached to an abelian
    surface}\label{table:genus2}
  \end{table}
\end{ex}

  %note: approximate moment computation suggests that rho is actually
  %irreducible over C. (same for genus 3 example)
  %this could be confirmed using the inner product technique
  %in the genus 2 example it is implied by Falting's theorem since End_Q(A) = Z

\begin{ex}
  Let~$A$ be the Jacobian of the genus~$3$ curve
  \lmfdbref{3.9-1.0.3-9-9.6}{HigherGenus/C/Aut/3.9-1.0.3-9-9.6} from the
  LMFDB~\cite{lmfdb:genus3}
  \[
    C : y^3 = x(x^3-1).
  \]
  Let~$K = \Q(\zeta_9)$.
  The curve~$C$ has an automorphism of order~$9$, defined over~$K$ and given
  by~$(x,y) \mapsto (\zeta_9^3 x, \zeta_9 y)$.
  In particular, the threefold~$A$ has CM by~$K$ defined over~$K$.
  %Let~$\ell = 19$, so that~$\ell$ is completely split in~$E$ and
  %therefore~$E_\ell = \Q_\ell$.
  %By point counting, we obtain that the Euler polynomial of~$A$ at~$p=5$
  %is
  %\[
  %  1+p^3T^6.
  %\]
  %The factorisation of this Euler polynomial over~$\F_\ell$ is
  %\[
  %  (T^2 + 4)(T^2 + 6)(T^2 + 9).
  %\]
  %In particular every invariant subspace of~$\rho_\ell$ has dimension~$0,2,4$
  %or~$8$.
  By point counting, the Euler polynomial of~$A$ at~$p=7$ is
  \[
    1 + pT^3 + p^3T^6,
  \]
  which is irreducible over~$\Q$, proving that~$A$ is simple.
  %and its factorisation over~$\F_\ell$ is
  %\[
  %  (T^3 + 9)(T^3 + 17),
  %\]
  %so that~$\rho_\ell$ is irreducible.
  Since~$K/\Q$ is Galois, $\pi(G)$ acts transitively on~$\Hom(K,\C)$ in the
  notations of Theorem~\ref{thm:cm}.
  There are two Galois orbits of CM types on~$K$: one lifted from the
  CM subfield~$\Q(\zeta_3)\subset K$, and a primitive one.
  Let~$\Phi^* = \{\zeta_9 \mapsto \exp(2i\pi \frac{4}{9}), \zeta_9 \mapsto
  \exp(2i\pi \frac{1}{9}), \zeta_9 \mapsto \exp(2i\pi \frac{2}{9}) \}$, which is
  primitive.
  By Theorem~\ref{thm:cm}, there
  exists an algebraic Hecke character~$\chi$ of~$K$ of type~$\Phi^*$ with values
  in~$K$ such that
  \[
    L(A,s) = L(\chi,s).
  \]
  Let~$\gp$ be the unique prime of~$K$ above~$3$.
  By computing resultants we see that~$A$ has good reduction away from~$3$.
  In particular the conductor of~$\chi$ is a power of~$\gp$, say~$\gp^m$.
  The restriction of~$\chi$ to~$(\Z_K/\gp^m)^\times$ has finite order and takes
  values in~$K$, and therefore has order dividing~$18$.
  By studying the $3$-adic convergence of~$(1+x)^{1/18}$ we see
  that~$1+\gp^{16}\subset (K_\gp^\times)^{18}$ and in particular we have~$m\le 16$.
  Alternatively, we could bound~$m$ by using the reduction theory of Picard
  curves~\cite{BouwPicard}, but the above method works in cases where no reduction theory
  is available.

  \begin{table}[h]
    \begin{tabular}[t]{c|c|c}
      prime~$\gr$ &  $\chi(\gr)\in \C$ & $\chi(\gr)\in K$ \\ \hline
      \texttt{3.1} & $\langle\exp(\frac{i\pi}{9})\rangle 1.73205\ldots i$
        & $\langle -\zeta_9\rangle\sqrt{-3} = \langle -\zeta_9\rangle(1+2\zeta_9^3) $ \\
      \texttt{19.1} & $4.34002\ldots + 0.40522\ldots i$
        & $2\zeta_9^5 + 2\zeta_9^4 + 2\zeta_9^3 + \zeta_9^2 - 2\zeta_9 + 2$ \\
      \texttt{19.2} & $-4.11721\ldots + 1.43128\ldots i$
        & $-\zeta_9^5 + 2\zeta_9^4 + 2\zeta_9^3 - 2\zeta_9^2 + 4\zeta_9 + 2$ \\
      \texttt{19.3} & $4.34002\ldots - 0.40522\ldots i$
        & $4\zeta_9^5 + \zeta_9^4 - 2\zeta_9^3 + 2\zeta_9^2 - \zeta_9$ \\
      \texttt{19.4} & $-4.11721\ldots - 1.43128\ldots i$
        & $-2\zeta_9^5 + \zeta_9^4 - 2\zeta_9^3 - 4\zeta_9^2 + 2\zeta_9$ \\
      \texttt{19.5} & $2.77718\ldots + 3.35964\ldots i$
        & $-\zeta_9^5 - 4\zeta_9^4 + 2\zeta_9^3 + \zeta_9^2 - 2\zeta_9 + 2$ \\
      \texttt{19.6} & $2.77718\ldots - 3.35964\ldots i$
        & $-2\zeta_9^5 - 2\zeta_9^4 - 2\zeta_9^3 + 2\zeta_9^2 - \zeta_9$ \\
      \texttt{37.1} & $4.34002\ldots - 4.26194\ldots i$
        & $4\zeta_9^5 + 4\zeta_9^4 - 2\zeta_9^3 + 5\zeta_9^2 + 2\zeta_9$ \\
      \texttt{37.2} & $2.77718\ldots - 5.41176\ldots i$
        & $-5\zeta_9^5 - 2\zeta_9^4 - 2\zeta_9^3 - \zeta_9^2 - 4\zeta_9$ \\
      \texttt{37.3} & $-4.11721\ldots - 4.47756\ldots i$
        & $-4\zeta_9^5 + 5\zeta_9^4 + 2\zeta_9^3 - 2\zeta_9^2 + 4\zeta_9 + 2$ \\
      \texttt{37.4} & $4.34002\ldots + 4.26194\ldots i$
        & $2\zeta_9^5 - \zeta_9^4 + 2\zeta_9^3 - 2\zeta_9^2 - 5\zeta_9 + 2$ \\
      \texttt{37.5} & $2.77718\ldots  + 5.41176\ldots i$
        & $2\zeta_9^5 - 4\zeta_9^4 + 2\zeta_9^3 + 4\zeta_9^2 + \zeta_9 + 2$ \\
      \texttt{37.6} & $-4.11721\ldots + 4.47756\ldots i$
        & $\zeta_9^5 - 2\zeta_9^4 - 2\zeta_9^3 - 4\zeta_9^2 + 2\zeta_9$ \\
      \texttt{64.1} & $-8$
        & $-8$
    \end{tabular}
    \caption{Values of the algebraic character~$\chi$ attached to an abelian
    threefold}\label{table:genus3}
  \end{table}

  Using our implementation we compute the group of characters of
  modulus~$\gp^{16}$. The subgroup of finite order characters is isomorphic
  to~$C_9^4$.
  There exists an algebraic character of type~$\Phi^*$, unique up to
  multiplication by a finite order character. Out of these $9^4 = 6561$
  candidate characters, checking that the value of~$a_{19}$ is sufficiently close to the
  value for~$A$, namely~$a_{19}(A) = 6$, eliminates all but~$2$ candidates.
  Checking that the value of~$a_{109}$ is sufficiently close to~$a_{109}(A) =
  -21$ leaves only one remaining candidate, which must therefore be~$\chi$.
  The conductor of~$\chi$ is~$\gp^4$ and~$\chi$ is in fact the unique algebraic
  character of type~$\Phi^*$ and conductor~$\gp^4$, and the restriction of~$\chi$
  to~$(\Z_K/\gp^4)^\times$ has order~$18$.
  The values of $\chi$ at some prime ideals \footnote{Labels are as
  in~\cite{labels} but with respect to the cyclotomic polynomial~$\Phi_9$, which
  is not the \texttt{polredabs} polynomial.} are given in
  Table~\ref{table:genus3}.
\end{ex}

\subsection{Density of gamma shifts}

The spectral parameters of an $L$-function are the gamma shifts
$\mu_j$ appearing in the gamma factor
\[
    \gamma(s) = \prod_{j=1}^{r_1} \Gamma_{\mathbb R}(s+\mu_j) \prod_{j={r_1+1}}^{r_1+r_2} \Gamma_{\mathbb C}(s+\mu_j).
\]
of its normalized functional equation $L(s)\gamma(s)=\Lambda(s)=\epsilon\overline{\Lambda}(1-s)$.
In this setting, the real parts $\Re(\mu_j)_{j\leq r_1}$ and $\Re(2\mu_j)_{j>r_1}$ are expected to be integers,
whereas the imaginary parts can be arbitrary transcendentals subject to
$\sum_{j=1}^{r_1} \mu_j+\sum_{j=r_1+1}^{r_1+r_2} 2\mu_j \in\R$.

As a matter of fact, Hecke characters allow us to attain a dense subspace of
these possible gamma shifts. The following statement must be well-known but we could
not find a reference for it.
\begin{prop}\label{prop:density}
    Let $r_1,r_2\geq 0$
    and $(\mu_j^*)\in(\set{0,1}+i\R)^{r_1}\times(\frac12\Z_{\geq0}+i\R)^{r_2}$
    a family of spectral parameters such that $\sum_{j\leq r_1} \mu_j^*+2\sum_{j>r_1} \mu_j^* \in\R$.

    Then for every number field $F$ of signature $(r_1,r_2)$ and every $\epsilon>0$,
    there exists a Hecke character $\chi$ of $F$ whose $L$-function gamma
    shifts~$\mu_j(\chi)$ satisfy
    \[
      |\mu_j(\chi) - \mu_j^*| < \epsilon.
    \]
\end{prop}
\begin{proof}
  Let~$F$ be a number field of signature $(r_1,r_2)$. For every modulus~$\gm$,
  let~$G_\gm \subset \widehat{F_\R^\times}$ be the image of the
  map~$\widehat{C}_\gm \to \widehat{F_\R^\times}$, that is, the group of
  infinity-types of characters of modulus~$\gm$.
  The group~$G_\gm$ is the group of elements~$\chi\in \widehat{F_\R^\times}$
  such that~$\chi(u)=1$ for all~$u\in \Z_F^\times(\gm) = \ker (\Z_F^\times \to
  (\Z_F/\gm)^\times)$.

  Let~$M>0$ be an integer.
  By the congruence subgroup property for unit groups of number
  fields~\cite[Th\'eor\`eme~$1$]{congrunits}, there exists a modulus~$\gm$ such
  that~$\Z_F^\times(\gm) \subset (\Z_F^\times)^M$.
  In particular, we get that
  \[
    \left\{\chi \in \widehat{F_\R^\times} \mid \chi^M \in G_1\right\} \subset G_\gm.
  \]
  Since the image of~$G_1$ in~$\R^{r_1+r_2}\times\Z^{r_2}$ has full rank, this
  proves that~$\bigcup_\gm G_\gm$ is dense in~$\widehat{F_\R^\times}$, which
  implies the claim.
\end{proof}

This makes Hecke characters good test cases for $L$-functions software, since
their coefficients are relatively easy to compute compared to other
transcendental automorphic forms.

\begin{ex}
  We exhibit a character of conductor $2^{20}$ over the real cubic field $F=\Q[x]/(x^3-3x+1)$
  whose parameters $\varphi_1$ and $\varphi_2$ approximate the constants $\pi$ and $e$ to 5 digits.
  %and parameters $\phi_1\approx3.141592$, $\phi_2\approx2.71828$, $\phi_3=-\phi_1-\phi_2$.
\begin{code}
> g=gcharinit(x^3-3*x+1,2^20); chi = [0,-2033118, 694865]~;
> gcharlocal(g,chi,1)
[0, 3.1415922385511383833775758885544915179]
> gcharlocal(g,chi,2)
[0, 2.7182831477529933175766620889117919084]
\end{code}
\end{ex}

\subsection{Partially algebraic Hecke characters}

In view of the special role played by algebraic Hecke characters, it is natural
to ask whether there exists partially algebraic Hecke characters, that is,
characters such that~$\varphi_\sigma=0$ for some~$\sigma$ but not
all \footnote{See \url{https://mathoverflow.net/questions/310706}}.
We provide a construction of such characters.

\begin{prop}~\label{prop:partcm}
  Assume~$F$ is a quadratic extension of another number field~$F_0$. Let~$R$ be
  the set of real places of~$F_0$ that become complex in~$F$, and let~$n_0$ be
  the degree of~$F_0$.
  Then for every modulus~$\gm$ of~$F$, there exists a subgroup~$H$
  of~$\widehat{C}_\gm$ of rank~$n_0$ in which every character
  satisfies~$\varphi_\sigma=0$ for every~$\sigma\in R$.
\end{prop}
\begin{proof}
  It suffices to prove the statement for the modulus~$\gm = 1$.
  Let~$g$ be the nontrivial element of~$\Gal(F/F_0)$, which acts
  on~$\widehat{C_1}$.
  Let~$H$ be the subgroup of~$\chi\in \widehat{C_1}$ such that there exists a
  finite order~$\xi\in \widehat{C_1}$ with~$\chi^g = \xi \chi^{-1}$.
  We have~$\rank(\widehat{C_1^1(F)})=n-1=2n_0-1$
  and~$\rank(\widehat{C_1(F_0)})=n_0-1$ (as is well-known but also easily seen
  from Proposition~\ref{prop:groupmodnorm}),
  so the rank of~$H$ is exactly~$n_0$.
  Moreover, for every infinite place~$\sigma$ of~$F$, every element of~$H$
  satisfies~$\varphi_{\sigma \circ g} = - \varphi_{\sigma}$.
  In particular for~$\sigma\in R$ this means that~$\varphi_\sigma = 0$.
\end{proof}

\begin{cor}
  Under the same hypotheses as Proposition~\ref{prop:partcm}, let~$r=0$ if~$F$
  does not contain a CM subfield and~$r$ be the degree of the maximal real
  subfield of~$F$ otherwise.
  Then for every modulus~$\gm$ of~$F$, there exists a subgroup~$H$
  of~$\widehat{C}_\gm$ of rank~$n_0-r$ in which every character
  satisfies~$\varphi_\sigma=0$ for every~$\sigma\in R$ and such that~$H$
  contains no nonzero almost-algebraic character.
  In particular, if~$F$ is not CM then there exists a partially algebraic
  character over~$F$.
\end{cor}
\begin{proof}
The integer~$r$ is the rank of the group of almost-algebraic characters.
\end{proof}

\begin{ex}
  Consider~$F_0 = \Q(\sqrt{5}) \subset F = \Q(5^{1/4})$.
\begin{code}
> gc=gcharinit(x^4-5,1);
> chi = [1,0,0]~;
> gcharlocal(gc,chi,1)
[0, -0.72908519629282042564585827345932876864]
> gcharlocal(gc,chi,2)
[0, 0.72908519629282042564585827345932876864]
> gcharlocal(gc,chi,3)
[2, 0]
\end{code}
  The character~$\chi$ satisfies %~$\varphi_{\sigma_1} = -0.729\dots$, $\varphi_{\sigma_2} = 0.729\dots$
  %and~$\varphi_{\sigma_3} = 0$,
  \[
    \chi_{\sigma_1} \colon x \mapsto |x|^{-i\times0.729\dots},\ 
    \chi_{\sigma_2} \colon x \mapsto |x|^{i\times0.729\dots},\text{ and }
    \chi_{\sigma_3} \colon z \mapsto (z/|z|)^{2},
  \]
  and is therefore an example of a partially
  algebraic character.
  Since~$n_0 = 2$ there is another independent partially algebraic character
  (namely~\texttt{[0,1,0]\~}).
\end{ex}

In a general number field~$F$, if one fixes a set of infinite places~$\Sigma$, a
natural question is to determine the group of \emph{$\Sigma$-algebraic}
characters, i.e. characters such that~$\varphi_\sigma = 0$ for
every~$\sigma\in\Sigma$.
The field~$F$ contains a maximal subfield~$K_0$ that is real at places
below~$\Sigma$, and may contain a quadratic extension~$K$ of~$K_0$ in which all
places below~$\Sigma$ are complex. When this is the case, one obtains a
corresponding group of $\Sigma$-algebraic characters. Does this construction
account for all the possible infinity types? Unlike the algebraic case where
Galois theory is sufficient to obtain a complete characterisation, the general case
seems to involve transcendence problems.

By automorphic induction to~$\GL_2$, partially algebraic characters yield
automorphic representations that are non-algebraic principal series at some
infinite places and discrete series at other ones. Analogously
to~\cite{partialwtone}, one may ask to explicitly construct such "partial Maass
forms" that do not come from Hecke characters. A possible way of doing this
would be to compute Maass forms on a well-chosen quaternion algebra and to use
the Jacquet--Langlands correspondence.

%induce to $\GL_2$ over a totally real field: constructs half-Maass half-discrete
%series automorphic representations.
%
%Refer to partial weight one
%\url{https://www.galoisrepresentations.com/2012/10/13/hilbert-modular-forms-of-partial-weight-one-part-i/}
%(mentions the CM ones, which we can also construct easily)
%and
%\url{https://www.galoisrepresentations.com/2012/12/08/hilbert-modular-forms-of-partial-weight-one-part-ii/}
%paper:
%\url{https://arxiv.org/abs/1407.3872}

\subsection{Twists and special values}

%Hecke characters is a natural twisting set for automorphic representations.
% conjecture: twists by enough to prove converse theorems?
Another interesting use of Hecke characters is to twist other $L$-functions to
obtain new ones. Our implementation makes it easy to follow
the experiments of \cite{Visser2021} on twists of elliptic curve $L$-functions.

Let $E/F$ be an elliptic curve of conductor $N_{E/F}$ over an imaginary quadratic field $F$,
and $\chi$ be an algebraic Hecke character of type $(a,b)$
and conductor $\gf$ over $F$.

Assume $\gcd(\gf,N_{E/F})=1$, then the twist
\[
  L(E\otimes\chi,s) = \sum_{(\gn,\gf)=1} a_\gn(E)\chi(\gn)\Nm(\gn)^{-s}
\]
conjecturally satisfies the functional equation
\[
  \Lambda(E\otimes \chi,s) = W \Lambda(E\otimes\overline{\chi},1+a+b-s)
\]
where
\[
  \Lambda(E\otimes \chi,s) =
  (\Nm(\gf)^2N_{E/F})^{\frac s2}
  \Gamma_\C(s-\min(a,b))
  \Gamma_\C(s-\min(a,b)-\mymathbb{1}_{a\neq b})
  L(E\otimes\chi,s)
\]
with special values predicted by Deligne's period conjecture~\cite{DelignePeriodes}.

\begin{ex}
    Let~$F=\Q(\sqrt{-43})$, $E/F$ the curve~\lmfdbref{43.1.a.1}{EllipticCurve/2.0.43.1/43.1/a/1}
    of equation $y^2+y=x^3+x^2$,
    and $\chi$ the algebraic character of conductor $1$ and type $(-2,2)$.

    We check numerically that the special value is a period related to $F$.
    \[
        \begin{aligned}
        L(E\otimes \chi,1)
        &\approx 2.996120826544463\dots \\
        &\approx \frac{2\pi}{\sqrt{43}^5} \Omega_F^8,
        \text{ where }
        \Omega_F=\sqrt{\prod_{a=1}^{42}\Gamma(\frac{a}{43})^{\left(\frac{-43}{a}\right)}}.
        \end{aligned}
    \]
\end{ex}

%TODO? no time
%Artin rep otimes Hecke char (irrep of Weil group but not of Galois group)

%Elliptic curve otimes Hecke char (irrep of Langlands group but not of Weil group
%or of Galois group)

%% Shimura over quadratic field

%%%Blasius, Harris, Lin Jie th 1.1
%%The special values of an $L$-function for a Hecke character over a CM field can be interpreted in terms of
%%CM periods.

%\bigskip $\ell$-adic algebraic Hecke characters modulo powers, and
%corresponding abelian extensions??
% Won't do (no time)

\ifarxiv
\else
\section{Data availability}
Data sharing is not applicable to this article as no datasets were generated or analysed during the current study.
\section{Statements and Declarations}
The authors certify that there is no actual or potential conflict of interest
in relation to this article.
\fi

\printbibliography

\newpage

\appendix

\section{Implementation notes}

This appendix collects notes on the matrix transformations used in our
implementation~\cite{parigp:gchar}.

Let~$\zeta$ be a generator of the group of roots of unity of~$F$,
let~$\Lambda_u$ (computed with~\texttt{bnfinit.fu}) be the image by~$\cL_S$ of
the span of a basis of~$\Z_F^\times/\langle \zeta\rangle$, and
let~$\Lambda_S$ (computed with~\texttt{bnfsunit[1]}) be the image by~$\cL_S$ of
the span of a basis of~$\Z_{F,S}^\times/\Z_F^\times$.
Let~$\Z_F^\times(\gm) = \ker(\Z_F^\times \to (\Z_F/\gm)^\times)$.
We define the following subgroup:
\[
  (\widehat{C}_\gm^1)_{k=0} = \left\{\chi\in \widehat{C}_\gm^1 \mid k_\sigma=0
  \text{ for every complex embedding }\sigma\right\}.
\]

We will describe a sequence of matrices representing a generating set
of~$\Lambda_0 =
\Lambda + \Z v_0 = \Lambda \stackrel{\perp}{\oplus} \Z v_0$. We only write the
ring to which the coefficients of the matrices belong.
We also indicate the number of rows and columns, with the following notations:
$n_s = |S|$, the integer~$n_c$ is the rank of~$\Lambda_\gm$ ($r(\gm)$ in the
main paper, but here we follow the notations from the code).

We want to apply matrix operations so that we compute a basis of~$\Lambda$ from
the generating set, exhibit interesting subgroups of the group of
Hecke characters, and preserve exactness of coefficients whenever possible.

At each step, we apply column operations to
modify the generating set of~$\Lambda$.
These column operations are obtained by applying a HNF reduction
to the submatrix displayed as a red block.
As a matter of fact, we rely on the following property:
from two lattices~$G,H$ the HNF computes a
subgroup~$H'$ of~$H$ that is saturated (i.e. the intersection of~$H$ with a
vector space), defined by some rows being~$0$, and a complement~$G'$ of~$H'$
in~$G+H$.

At the end we compute an inverse to get a basis of~$\Lambda_0^\vee$ from which
we deduce a basis of~$\Lambda^\vee$, and we describe
various subgroups of the group of Hecke characters that appear naturally.
In the end tables, the meaning of the rows is as follows: the title of the
row is the subgroup generated by all the previous rows.
In other words, the corresponding rows generate a complement of the previous
rows in the subgroup in the title of the row.
The column labelled~$\chi(\gp)$ contains the
values~$\frac{1}{2\pi}\arg\chi(\gp)$ and the column labelled~$\chi_\gm(g_i)$
contains the values~$\frac{1}{2\pi}\arg\chi(\log_\gm^{-1}(g_i))$.

\subsection{Case without a CM subfield}

Here we do not assume that we have a CM subfield. This subsection is no longer
implemented, but serves as a simpler version of the next subsection.

Initial matrix.
\[
  \begin{array}{cc|ccccc}
            &            & n_s       & r_1+r_2-1 & n_c       & 1     & r_2      \\
            &            & \Lambda_S & \Lambda_u & \Lambda_{\gm}    & \zeta & \Z^{r_2} \\ \hline
    n_s     & v_S         & \Z        & 0         & 0         & 0     & 0 \\
    n_c     & \log_\gm    & \Z        & \Z        & \hi\Z     & \hi\Z & 0 \\
    r_1+r_2 & \log_\sigma & \R        & \R        & 0         & 0     & 0 \\
    r_2     & \arg_\sigma & \R        & \R        & 0         & \Q    & \Z
  \end{array}
\]

\begin{enumerate}
    \renewcommand{\labelenumi}{Step \arabic{enumi}. }
\item
    We compute the subgroup~$\langle\zeta(\gm)\rangle =
    \langle\zeta\rangle\cap\Z_F^\times(\gm)$ and a
    complement~$\Lambda_{\gm,\zeta}$ of~$\langle\zeta(\gm)\rangle$ in~$\Lambda_{\gm} +
    \langle\zeta\rangle$.

\[
  \begin{array}{cc|ccccc}
            &             & n_s       & r_1+r_2-1 & n_c           & 1          & r_2 \\
            &             & \Lambda_S & \Lambda_u & \Lambda_{\gm,\zeta} & \zeta(\gm)  & \Z^{r_2} \\ \hline
    n_s     & v_S         & \Z        & 0         & 0             & 0          & 0 \\
    n_c     & \log_\gm    & \Z        & \hi\Z     & \hi\Z         & 0          & 0 \\
    r_1+r_2 & \log_\sigma & \R        & \R        & 0             & 0          & 0 \\
    r_2     & \arg_\sigma & \R        & \R        & \Q            & \Q         & \Z
  \end{array}
\]

\item
  We compute the span~$\Lambda_u(\gm)\subset\Lambda_u$ of a basis
    of~$\Z_F^\times(\gm)/\langle\zeta(\gm)\rangle$ and a
    complement~$\Lambda_{\gm,u}$ of~$\Lambda_u(\gm)$ in~$\Lambda_{\gm,\zeta} +
    \Lambda_u$.
%We also swap~$v_0$ and~$\zeta(\gm)$.

\[
  \begin{array}{cc|ccccc}
            &             & n_s      & n_c       & r_1+r_2-1    & 1         & r_2      \\
            &             & \Lambda_S & \Lambda_{\gm,u}  & \Lambda_u(\gm)  & \zeta(\gm) & \Z^{r_2} \\ \hline
    n_s     & v_S         & \Z        & 0         & 0            & 0         & 0 \\
    n_c     & \log_\gm    & \Z        & \Z        & 0            & 0         & 0 \\
    r_1+r_2 & \log_\sigma & \R        & \R        & \R           & 0         & 0 \\
    r_2     & \arg_\sigma & \R        & \R        & \R           & \hi\Q     & \hi\Z
  \end{array}
\]

\item
We compute~$\Lambda_{\zeta} = \Z^{r_2}+\langle\zeta(\gm)\rangle$.

\[
  \begin{array}{cc|cccc}
            &             & n_s      & n_c       & r_1+r_2-1    & r_2      \\
            &             & \Lambda_S & \Lambda_{\gm,u}  & \Lambda_u(\gm)  & \Lambda_\zeta \\ \hline
    n_s     & v_S         & \Z        & 0         & 0            & 0         \\
    n_c     & \log_\gm    & \Z        & \Z        & 0            & 0         \\
    r_1+r_2 & \log_\sigma & \R        & \R        & \R           & 0         \\
    r_2     & \arg_\sigma & \R        & \R        & \R           & \Q
  \end{array}
\]

\item
    We include $v_0$ to obtain a square block on the $\log_\sigma$ components.

\[
  \begin{array}{cc|ccccc}
            &             & n_s       & n_c       & r_1+r_2-1     & 1     & r_2      \\
            &             & \Lambda_S & \Lambda_{\gm,u}  & \Lambda_u(\gm)   & v_0   & \Lambda_\zeta \\ \hline
    n_s     & v_S         & \Z        & 0         & 0             & 0     & 0         \\
    n_c     & \log_\gm    & \Z        & \Z        & 0             & 0     & 0         \\
    r_1+r_2 & \log_\sigma & \R        & \R        & \R            & \Z    & 0         \\
    r_2     & \arg_\sigma & \R        & \R        & \R            & 0     & \Q
  \end{array}
\]

    \item
We now compute the dual lattice by taking the inverse of the matrix (note that all the diagonal blocks are invertible).
%The transpose is as follows.
%\[
%  \begin{array}{c|cccc}
%                   & v_S   & \log_\gm  & \log_\sigma   & \arg_\sigma \\ \hline
%    \Lambda_S      & \Z    & \Z        & \R            & \R \\
%    \Lambda_{\gm,u}       & 0     & \Z        & \R            & \R \\
%    \Lambda_u(\gm) & 0     & 0         & \R            & \R \\
%    v_0            & 0     & 0         & \Z            & 0 \\
%   \Lambda_\zeta    & 0     & 0         & 0             & \Q
%  \end{array}
%\]
We
obtain the following shape, where the red blocks a priori have coefficients
in~$\R$. However, the coefficients in the~$\chi_\gm(g_i)$ columns (dual to the
previous~$\log_\gm$ rows) represent a character
on~$(\Z_F/\gm)^\times$, which is a finite group, so they must be rationals with
denominator divisible by the exponent of this group, and the coefficients in
the~$k_\sigma$ columns (dual to the previous~$\arg_\sigma$ rows) represent
characters on~$(\R/\Z)^{r_2}$, so they must be integers.

%\[
%  \begin{array}{c|ccccc}
%                & \widehat{\Cl_F}   & \widehat{\Cl(\gm)}    & (\widehat{C}_\gm^1)_{k=0} & v_0^\vee  &  \widehat{C}_\gm^1 \\ \hline
%    v_S         & \Q                & \Q                    & \R                        & \R        & \R \\
%    \log_\gm    & 0                 & \Q                    & \rr\Q                     & \rr\Q     & \rr\Q \\
%    \log_\sigma & 0                 & 0                     & \R                        & \R        & \R \\
%    \arg_\sigma & 0                 & 0                     & 0                         & 0         & \rr\Z
%  \end{array}
%\]

\[
  \begin{array}{cc|cccc}
              &                             & n_s           & n_c           & r_1+r_2           & r_2 \\
              &                             & \chi(\gp)     & \chi_\gm(g_i) & \varphi_\sigma    & k_\sigma \\ \hline
    n_s       & \widehat{\Cl_F}             & \Q            & 0             & 0                 & 0 \\
    n_c       & \widehat{\Cl(\gm)}          & \Q            & \Q            & 0                 & 0 \\
    r_1+r_2-1 & (\widehat{C}_\gm^1)_{k=0}   & \R            & \rr\Q         & \R                & 0 \\
    1         & v_0^\vee                    & \R            & \rr\Q         & \R                & 0 \\
    r_2       & \widehat{C}_\gm^1           & \R            & \rr\Q         & \R                & \rr\Z
  \end{array}
\]

\item
We remove~$v_0^\vee$, obtaining a basis of~$\Lambda^\vee$.

%\[
%  \begin{array}{c|cccc}
%                & \widehat{\Cl_F}   & \widehat{\Cl(\gm)}    & (\widehat{C}_\gm^1)_{k=0} & \widehat{C}_\gm^1 \\ \hline
%    v_S         & \Q                & \Q                    & \R                        & \R \\
%    \log_\gm    & 0                 & \Q                    & \Q                        & \Q \\
%    \log_\sigma & 0                 & 0                     & \R                        & \R \\
%    \arg_\sigma & 0                 & 0                     & 0                         & \Z
%  \end{array}
%\]

\[
  \begin{array}{cc|cccc}
              &                             & n_s           & n_c           & r_1+r_2           & r_2 \\
              &                             & \chi(\gp)     & \chi_\gm(g_i) & \varphi_\sigma    & k_\sigma \\ \hline
    n_s       & \widehat{\Cl_F}             & \Q            & 0             & 0                 & 0 \\
    n_c       & \widehat{\Cl(\gm)}          & \Q            & \Q            & 0                 & 0 \\
    r_1+r_2-1 & (\widehat{C}_\gm^1)_{k=0}   & \R            & \Q            & \R                & 0 \\
    r_2       & \widehat{C}_\gm^1           & \R            & \Q            & \R                & \Z
  \end{array}
\]

\end{enumerate}

\subsection{Case with a CM subfield}

In this section we assume that~$F$ contains a CM subfield.
The implementation takes advantage of the following rationality result.
\begin{lem}
  \label{lem:int-narg}
  Let~$K$ be a CM field and let~$\rootsofone{K}$ be the number of roots of unity in~$K$. Then for
  every~$u\in\Z_K^\times$ and~$\sigma\colon K\hookrightarrow\C$, we have
  \[
    \frac{\arg \sigma(u)}{2\pi} \in \frac{1}{2\rootsofone{K}}\Z.
  \]
\end{lem}
\begin{proof}
  Let~$z = u/\bar{u}\in\Z_K^\times$. Then for every complex embedding~$\sigma$,
  we have~$|\sigma(z)|=1$. So~$z$ is a root of unity: $z^{\rootsofone{K}}=1$.
  We obtain
  \[
    2\arg(\sigma(u)) = \arg(z) \in \frac{2\pi}{\rootsofone{K}}\Z,
  \]
  hence the result.
\end{proof}

For the remainder of this section, we assume that we are given~$K$ the maximal
CM subfield of~$F$. We will write~$\tau$ for the complex embeddings of~$K$ and~$\sigma$ for
the complex embeddings of~$F$. For every complex embedding~$\tau$ of~$K$, we
let~$N\tau(x) = \prod_{\sigma\mid\tau}\sigma(x)$.
%and we fix once and for all an
%extension~$\tilde{\tau}$ of~$\tau$ to~$F$.

We start by applying Steps 1 to 3 as in the previous case,
obtaining the same shape, but we change the order of the columns.

\[
  \begin{array}{cc|ccccc}
            &             & n_s       & n_c               & r_2           & r_1+r_2-1 \\
            &             & \Lambda_S & \Lambda_{\gm,u}   & \Lambda_\zeta & \Lambda_u(\gm)  \\ \hline
    n_s     & v_S         & \Z        & 0                 & 0             & 0            \\
    n_c     & \log_\gm    & \Z        & \Z                & 0             & 0            \\
    r_1+r_2 & \log_\sigma & \R        & \R                & 0             & \R           \\
    r_2     & \arg_\sigma & \R        & \R                & \Q            & \R
  \end{array}
\]

We now focus on the archimedean block, where we will apply extra column
operations to exhibit the subgroup of almost-algebraic characters.

\begin{enumerate}
\renewcommand{\labelenumi}{Step \arabic{enumi}'. }
\setcounter{enumi}{3}
\item
We introduce extra rows, parametrised by the complex embeddings~$\tau$ of~$K$,
with values~$\arg\bigl(N\tau(\epsilon)\bigr)/2\pi$
for~$\epsilon\in
    \Lambda_u(\gm)$. Those values are in~$\frac{1}{2\rootsofone{K}}\Z$ by Lemma~\ref{lem:int-narg}.
    We also select a subset of~$r_2-r_2(K)$ complex embeddings such that the corresponding
coordinates on~$\R^{r_2}$ are linearly independent, as linear forms, of the ones
corresponding to~$N\tau$. In the following matrices, we label the corresponding
row by~$\arg'$.

\[
  \begin{array}{cc|cc}
                &                   & r_2              & r_1+r_2-1 \\
                &                   & \Lambda_\zeta    & \Lambda_u(\gm)  \\ \hline
    r_1+r_2     & \log_\sigma       & 0                & \R           \\
    r_2-r_2(K)  & \arg'             & \Q               & \R           \\
    r_2(K)      & \arg_{N\tau}      & \hi\Q            & \hi\Q
  \end{array}
\]

\item
We apply a column HNF on the red blocks: this computes the
    subgroup~$\Lambda(\arg)$ of elements of~$\Lambda_{\zeta} + \Lambda_u(\gm)$
that have trivial~$\arg N\tau$, and a complement~$\Lambda_{\arg}$
    of~$\Lambda(\arg)$ in~$\Lambda_{\zeta}+\Lambda_u(\gm)$.
%The rank of~$\Lambda(\arg)$ is~$n-1+r_2(K)$, and the rank of~$\Lambda_{\arg}$ is~$r_2(K)$.
We get the following shape.

\[
  \begin{array}{cc|cc}
                &                 & r_2(K)          & n-r_2(K)-1 \\
                &                 & \Lambda_{\arg}  & \Lambda(\arg)    \\ \hline
    r_1+r_2     & \log_\sigma     & \R              & \R          \\
    r_2-r_2(K)  & \arg'           & \R              & \R          \\
    r_2(K)      & \arg_{N\tau}    & \Q              & 0
  \end{array}
\]

\item
  We insert~$v_0$ as before.

\[
  \begin{array}{cc|ccc}
                &                 & r_2(K)          & n-r_2(K)-1    & 1 \\
                &                 & \Lambda_{\arg}  & \Lambda(\arg) & v_0   \\ \hline
    r_1+r_2     & \log_\sigma     & \R              & \R            & \Z \\
    r_2-r_2(K)  & \arg'           & \R              & \R            & 0 \\
    r_2(K)      & \arg_{N\tau}    & \Q              & 0             & 0
  \end{array}
\]

  Note that the block corresponding to columns~$\Lambda(\arg)$ and~$v_0$ and
  rows~$\log_\sigma$ and~$\arg'$ is square and invertible.

\item
  Now we compute, as before, the dual basis of~$\Lambda^\vee$ by taking the
        inverse matrix.
  %On the archimedean block, the transposed matrix is as follows.
  %\[
  %  \begin{array}{c|cccc}
  %                        & \log_\sigma   & \arg'  & \arg_{N\tau}  \\ \hline
  %    \Lambda_{\arg}            & \R            & \R     & \Q            \\
  %    \Lambda(\arg)            & \R            & \R     & 0             \\
  %    v_0                 & \Z            & 0      & 0
  %  \end{array}
  %\]

  %The inverse of this matrix is a priori of the following shape.
  On the archimedean block, the inverse of the matrix is a priori of the
  following shape.
  %\[
  %  \begin{array}{c|cccc}
  %                     &            &           & v_0^\vee  \\ \hline
  %    \varphi_{\sigma} & 0          & \R        & \R    \\
  %    k'               & 0          & \R        & \R    \\
  %    k_{N\tau}        & \Q         & \R        & \R
  %  \end{array}
  %\]
  \[
    \begin{array}{cc|ccc}
                 &             & r_1+r_2           & r_2-r_2(K) & r_2(K) \\
                 &             & \varphi_{\sigma}  & k_\sigma'  & k_{N\tau} \\ \hline
      r_2(K)     & \text{a.a.} & 0                 & 0          & \Q \\
      n-r_2(K)-1 &             & \R                & \R         & \R \\
      1          & v_0^\vee    & \R                & \R         & \R
    \end{array}
  \]

\item
  We delete~$v_0^\vee$ and change coordinates again to recover the usual parameters~$k_\sigma$.
  As in Step 5 above, the red blocks are a priori real but they must actually be
  integers.
  %\[
  %  \begin{array}{c|cccc}
  %                     &            &         \\ \hline
  %    \varphi_{\sigma} & 0          & \R      \\
  %    k_\sigma         & \rr{\Z}    & \rr{\Z}
  %  \end{array}
  %\]

  \[
    \begin{array}{cc|cc}
                 &               & r_1+r_2           & r_2 \\
                 &               & \varphi_{\sigma}  & k_\sigma \\ \hline
      r_2(K)     & \text{a.a.}   & 0                 & \rr\Z \\
      n-r_2(K)-1 &               & \R                & \rr\Z
    \end{array}
  \]

In the end, we obtain the following shape for the matrix of characters.

%\[
%  \begin{array}{c|cccc}
%                & \widehat{\Cl_F}   & \widehat{\Cl(\gm)}    & \Cmalg     & \widehat{C}_\gm^1 \\ \hline
%    v_S         & \Q                & \Q                    & \R         & \R \\
%    \log_\gm    & 0                 & \Q                    & \Q         & \Q \\
%    \log_\sigma & 0                 & 0                     & 0          & \R \\
%    \arg_\sigma & 0                 & 0                     & \Z         & \Z
%  \end{array}
%\]

\[
  \begin{array}{cc|cccc}
               &                     & n_s       & n_c           & r_1+r_2           & r_2 \\
               &                     & \chi(\gp) & \chi_\gm(g_i) & \varphi_\sigma    & k_\sigma \\ \hline
    n_s        & \widehat{\Cl_F}     & \Q        & 0             & 0                 & 0 \\
    n_c        & \widehat{\Cl(\gm)}  & \Q        & \Q            & 0                 & 0 \\
    r_2(K)     & \Cmalg              & \R        & \Q            & 0                 & \Z \\
    n-r_2(K)-1 & \widehat{C}_\gm^1   & \R        & \Q            & \R                & \Z
  \end{array}
\]

This matrix is accessible as \texttt{gcharinit(bnf,mod)[1]} in our
implementation in Pari/GP 2.15~\cite{parigp:gchar}.
As explained in section~\ref{sec:almostalg}, we can recover the group of
algebraic Hecke characters from~$\Cmalg$.

\end{enumerate}

\end{document}